\documentclass[preprint,12pt,authoryear]{elsarticle}
\usepackage{amsmath,txfonts}
\usepackage{graphicx}

\usepackage{url}
\usepackage{subcaption}
\usepackage[colorlinks=true,linkcolor=black, citecolor=blue, urlcolor=blue]{hyperref}

 \usepackage{graphicx,rotating,booktabs}
\usepackage{float}

\usepackage[verbose]{placeins}
\usepackage[ngerman, english]{babel}
\usepackage{blindtext}

\setcitestyle{comma,numbers,super}
\usepackage{graphicx}

\usepackage{amssymb}

\journal{Statistics in Medicine}

\begin{document}

\begin{frontmatter}
\title{Addressing Spatially Structured Interference in Causal Analysis Using Propensity Scores}

\newcommand{\textunderscript}[1]{$_{\text{#1}}$}

\def\independenT#1#2{\mathrel{\rlap{$#1#2$}\mkern2mu{#1#2}}}

\author[label1]{Keith W. Zirkle}
\author[label2]{Marie-Ab\`ele Bind}
\author[label3]{Jenise L. Swall}
\author[label1]{David C. Wheeler}

\address[label1]{Department of Biostatistics, Virginia Commonwealth University, Richmond, VA, USA}
\address[label2]{Department of Statistics, Harvard University, Cambridge, MA, USA}
\address[label3]{Department of Statistical Sciences and Operations Research, Virginia Commonwealth University, Richmond, VA, USA}

\begin{abstract}
\indent Environmental epidemiologists are increasingly interested in establishing causality between exposures and health outcomes. A popular model for causal inference is the Rubin Causal Model (RCM), which typically seeks to estimate the average difference in study units' potential outcomes. An important assumption under RCM is no interference; that is, the potential outcomes of one unit are not affected by the exposure status of other units. The no interference assumption is violated if we expect spillover or diffusion of exposure effects based on units' proximity to other units and several other causal estimands arise. Air pollution epidemiology typically violates this assumption when we expect upwind events to affect downwind or nearby locations. This paper adapts causal assumptions from social network research to address interference and allow estimation of both direct and spillover causal effects. We use propensity score-based methods to estimate these effects when considering the effects of the Environmental Protection Agency's 2005 nonattainment designations for particulate matter with aerodynamic diameter $<2.5 \mu m$ (PM$_{2.5}$) on lung cancer incidence using county-level data obtained from the Surveillance, Epidemiology, and End Results (SEER) Program. We compare these methods in a rigorous simulation study that considers both spatially autocorrelated variables, interference, and missing confounders. We find that pruning and matching based on the propensity score produces the highest probability coverage of the true causal effects and lower mean squared error. When applied to the research question, we found protective direct and spillover causal effects.

\end{abstract}


\end{frontmatter}


{\em Research reported in this publication was supported by the National Institute of Environmental Health Sciences of the National Institutes of Health under Award Number T32ES007334, by VCU Massey Cancer Center's Cancer Prevention and Control 2018 Scholarship, by the Office of the Director, National Institute of Health under Award Number DP5OD021412, and by the John Harvard Distinguished Science Fellows Program, within the FAS Division of Science of Harvard University. The content is solely the responsibility of the authors and does not necessarily represent the official views of the National Institutes of Health.}

\section{Introduction}
\label{S:1}
Many public health studies aim to estimate causal relationships between some exposure or treatment and health-related outcomes; however, most epidemiologic studies are observational.
Observational data present a distinct, but well-studied challenge to address causality. The ideal paradigm for causal analysis is randomization that, on average, leads to similar treated and non-treated subjects, and the treatment effect can be estimated unbiasedly. In observational studies, the treatment or exposure is not typically randomized and must be treated as a conditionally randomized experiment\citep{Hernan06}. Generally, the treatment assignment mechanism is characterized to emulate randomization and usually involves adjusting for multiple covariates known to affect the observed treatment assignment. In other words, the treatment assignment $Z$ should be independent of the potential outcomes $Y(Z)$ given covariates $X$. This is called ignorability and is expressed as
\begin{equation}\label{ignorability}
P(Z|Y(0),Y(1),X)=P(Z|X).
\end{equation}

Rubin's Causal Model (RCM) is the most popular paradigm for defining causal effects\citep{Rubin74}. Several popular methods exist that utilize RCM, including balancing scores, outcome regression, and principal stratification. For every unit, the potential outcomes are considered, i.e. the unit's outcome under exposure to the treatment and the unit's outcome not under exposure to the treatment. The difference in these potential outcomes is considered a causal effect where all other things are held constant except for exposure to the treatment. RCM relies on several assumptions whose validity varies study to study, including the stable unit treatment value assumption (SUTVA), first proposed by Cox in 1958\cite{Cox58,RubinDonaldB.1984BJaR}. SUTVA states that (i) there are not multiple forms of a treatment (called ``consistency") and (ii) a unit's exposure or treatment does not affect the outcome of other units (called ``no interference"). Until recently, causal inference methods either struggled to or introduced restrictive assumptions to handle data that may have interference.

We typically expect interference in studies when the outcome or exposure is related to other observations. Readily available examples of interference include infectious diseases \cite{Halloran95}, social networks \citep{Kao17,Toulis12}, educational programs \citep{Hong06}, crime prevention strategies \cite{Verbitsky12}, and air pollution \cite{Zigler12}. This interference cannot be ignored without making erroneous inferences \citep{VanderWeele12,Sobel06}. More often, the interference is more than a nuisance. Different causal estimands arise under interference, including a direct effect and an indirect (spillover) effect. This spillover effect between units or groups of units can be of research interest \citep{Hudgens08}.

In the last decade, causal methods that handle interference have grown \citep{Hong06,Verbitsky12,Zigler12,VanderWeele12}.
Most methods make assumptions about the interference structure to allow for relevant estimands. These assumptions are not always appropriate for our application. Hong and Raudenbush proposed unit-specific interference as low or high; they assumed potential outcomes are a function of a subject's treatment status and the interference status\citep{Hong06}. Sobel proposed a partial interference assumption where subjects are grouped into classes and no interference is assumed between subjects in different classes\citep{Sobel06}. Zigler et al. further extended Sobel's partial interference assumption into an assignment group interference assumption (AGIA) where locations within a regulated area do not affect locations in an unregulated area\cite{Zigler12}. They found AGIA may not hold if interference is expected to occur between exposed and control units. Random effects with spatial structure have also been used to account for interference \citep{Zigler17}, but may be insufficient for capturing spillover between subjects.

We motivate the methods presented in this paper by considering air pollution studies. Spatial interference exists in air pollution applications by air pollution's very nature. We expect upwind events to affect downwind locations and, thus, expect interference to be directional based on wind patterns. We expect to find a direct causal effect of air pollution (or air pollution regulation) in an area exposed directly to treatment (e.g. regulation) and may expect a spillover effect in downwind areas. Researchers are increasingly interested in these effects because they allow for more strategic initiatives and program implementations and also may provide stronger evidence for regulatory control\cite{Zigler14}.

In this paper, we propose methods for estimating air pollution interference by elucidating spatial relationships and adapting certain causal assumptions to a spatial setting where there is expected spillover of an exposure. We characterize both a treatment assignment and interference mechanism and adjust for confounders for both using propensity score modeling. Our assumptions differ from previous work in air pollution epidemiology, specifically Zigler et al., by allowing interference to occur between exposed and control units and also adjusting for confounding of the interference mechanism\cite{Zigler12}. To evaluate our approach, we compare the methods when applied to simulated datasets where we introduce both spatially autocorrelated covariates and treatment assignment. We finally apply the methods to an air pollution dataset.

\section{Methods}
\subsection{Notation and Causal Estimands}

\begin{figure}[h!]
 \includegraphics[width=\textwidth]{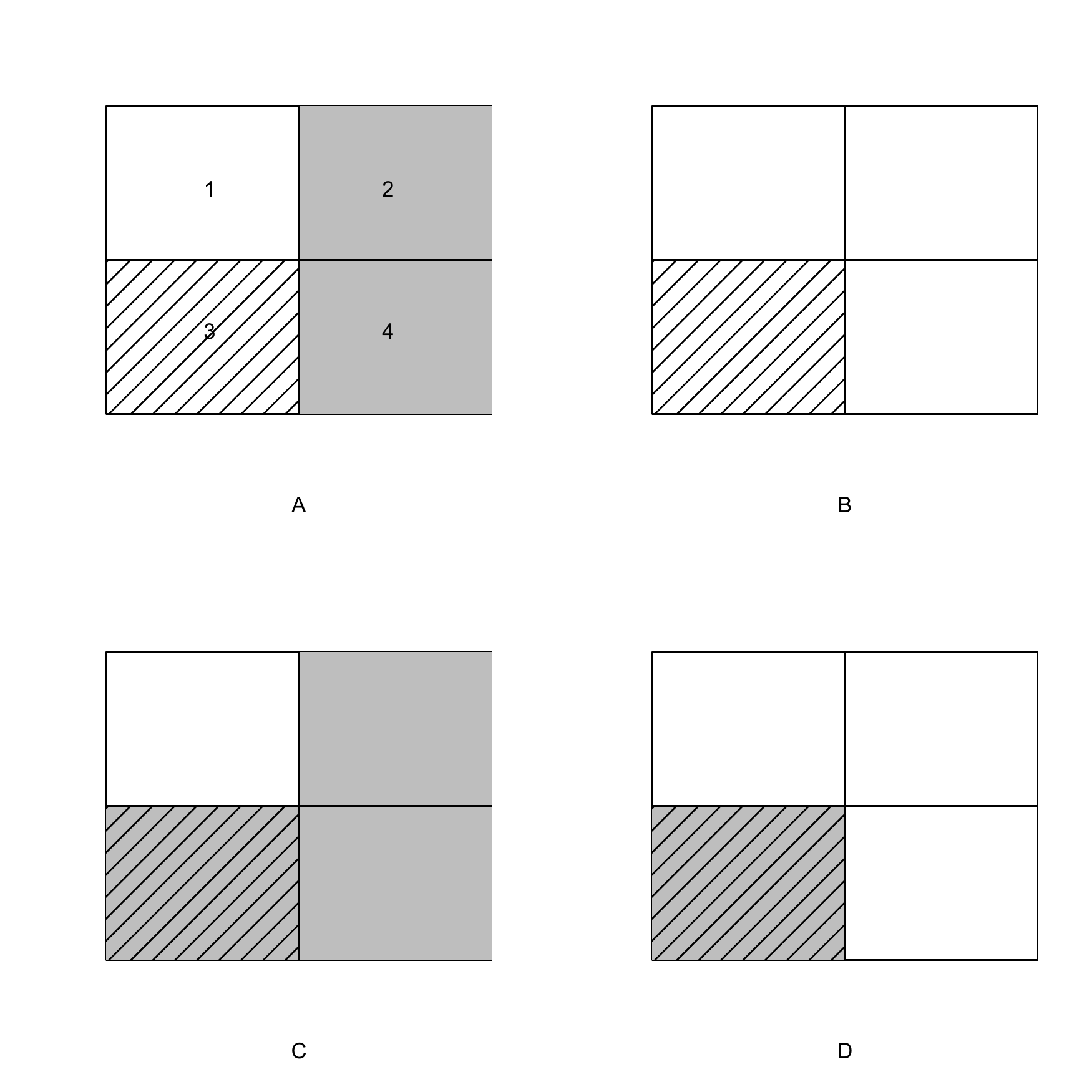}
\caption[Example Toy Figure.]{An example study population of $N=4$ areas where diagonally shaded areas denote if the area received treatment. Consider area $i=3$ (marked with diagonal lines). For area $i=3$, we define the individual direct causal effect to be the contrast in outcomes between panels A and C Similarly, for area $i=3$, we define individual indirect causal effects to be the contrast in outcomes between panels A and B and between panels C and D. The difference between these two individual indirect causal effects is the individual treatment status $Z$ ($Z_3=0$ in panels A and B and $Z_3=1$ in panels C and D).}
\label{fig:ToyExample}
\end{figure}

Consider a study population of $i=1,\ldots,N$ contiguous areal units at a certain time. Let $Y_i$ be a count outcome such as number of deaths per unit. Denote treatment $Z_i\in \{0,1\}$ where $Z_i=1$ if unit $i$ receives the treatment and $Z_i=0$ otherwise, and let $\mathbf{Z}$ denote the treatment allocation for all units. We illustrate these concepts in Figure~\ref{fig:ToyExample} for a population of $N=4$ areas (panel A). In Figure~\ref{fig:ToyExample}, we shade treated areas and can express $\mathbf{Z}=\{0,1,0,1\}$ for the study area represented in panel A. Under no interference, each unit has two potential outcomes $Y_i(Z_i)$ expressed as $[Y_i(1),Y_i(0)]$. We only ever observe one of these outcomes dependent on the observed value of $Z_i$. We define a causal effect as $Y_i(1)-Y_i(0)$. In Figure~\ref{fig:ToyExample}, this would be akin to finding the difference in outcomes for area $i=3$ in panels A and C. 

With general interference, the potential outcomes become $Y_i(\mathbf{Z})=Y_i(Z_i,\mathbf{Z}_{-i})$ where $Z_i$ is the individual treatment for unit $i$ and $\mathbf{Z}_{-i}$ is the treatment vector for all other units in the population. We may reasonably expect interference is limited to a spatial neighborhood structure. We denote this structure through an $N\times N$ network matrix $\mathbf{A}$ where element $A_{ij}>0$ if a relationship exists between units $i$ and $j$. We note that $A_{ij}$ may not equal $A_{ji}$ if the relationship is directional. For example, we expect air pollution interference to vary with or to be a function of wind direction so that pollution in unit $i$ blows downwind to unit $j$, but not vice versa.

Let $\mathbb{N}_i$ represent all units that may affect unit $i$, which we call a neighborhood. While other units in a population may affect unit $i$, this is computationally intensive and challenging to interpret so we often restrict analysis to only first-order, or also called adjacent, neighbors. Let $\mathbb{N}_i$ represent the neighborhood, and let $\mathbf{A}$ represent a network matrix for these neighbors. Then $A_{ij}>0$ if units $i$ and $j$ are first-order neighbors, or immediately adjacent. Denote the size of the neighborhood of unit $i$ as $m_i=\sum_{j=1}^nA_{ij}=|\mathbb{N}_i|$.

In Figure~\ref{fig:ToyExample}, each area has three first-order neighbors: $\mathbb{N}_{i=1}$ contains areas 2, 3, and 4, $\mathbb{N}_{i=2}$ contains units 1, 3, and 4 as first-order neighbors, etc.  We may express $\mathbf{A}=\mathbf{I}_{4\times4}$ because every area is adjacent with each other.
For all the areas, $m_i=3$; that is, each area has three neighbors.

For interference, we are interested in the treatment allocation of $\mathbb{N}_i$, $\mathbf{Z}_{\mathbb{N}_i}=[z_j]_{j\in \mathbb{N}_i}\in\{0,1\}^{m_i}$. The set of treated units in neighborhood $\mathbb{N}_i$ are $\mathbb{N}_i^{\mathbf{z}}$, and we consider interference to be a function of $\mathbb{N}_i^{\mathbf{z}}$. We will call this a functional neighborhood interference assumption, which is similar to assumptions proposed by several others \cite{Verbitsky12, Sussman17,Hong06,Athey15,Manski13}. We define our potential outcomes as
\begin{equation}\label{eq:po}
Y_i(Z_i,f(\mathbf{Z}_{\mathbb{N}_i})).\tag{Definition 1}
\end{equation}
In other words, a unit-level potential outcome is dependent both on its own treatment assignment and some function of the treatment assigned to its first-order neighbors. In Figure~\ref{fig:ToyExample}, we may write the potential outcomes for area 3 in panel A as $Y(Z=0,f(\{0,1,1\}))$ where $\{0,1,1\}$ represents the treatment allocation of areas 1, 2, and 4 (the neighbors of area 3).

Until this point, we assume that interference occurs between units $i$ and $j$ based on proximity, e.g. neighborhood structure. We also suggest that the relationship can be directional, e.g. based on wind direction in air pollution studies. From this, we argue that there exists both a treatment assignment mechanism and an interference mechanism, which can be characterized as the network $\mathbf{A}$. Defining an interference mechanism is not well established in the literature and will be contextual to the data and the expected ``spillover" of a treatment or exposure. Information on the spatial neighborhood structure may inform this network as well as the observed outcomes \cite{Kao17}.

Using typical causal inference estimand language where we estimate effects either from randomized experiment data or meet other causal assumptions necessary for causal analysis with observational data, we define a direct causal effect as:
 \begin{equation}\label{eq:de_i}
DE_i(\mathbf{z})=Y_i(Z_i=1,f(\mathbf{Z}_{\mathbb{N}_i}=\mathbf{z}))-Y_i(Z_i=0,f(\mathbf{Z}_{\mathbb{N}_i}=\mathbf{z}))
\end{equation}
where $\mathbf{z}$ represents an observed treatment vector. Kao termed this a ``primary" treatment causal effect in social influence networks\cite{Kao17} . In Figure~\ref{fig:ToyExample}, we would compare the outcomes for area 3 in panels A and C to estimate an individual direct causal effect. The average direct causal effect can be defined as:
\begin{equation}\label{eq:de}
\overline{DE}=\frac{1}{N} \sum_{i=1}^N DE_i(\mathbf{z}).
\end{equation}

We define an indirect causal effect as:
\begin{equation}\label{eq:ie_i}
IE_{i}(z)=Y_i(Z_i=z,f(\mathbf{Z}_{\mathbb{N}_i}=\mathbf{z}))-Y_i(Z_i=z,f(\mathbf{Z}_{\mathbb{N}_i}=\mathbf{0})).
\end{equation}
This indirect effect is defined based on which $k$th-order neighbors are assumed to influence unit $i$. First-order neighbors are areas immediately adjacent to area $i$, second-order neighbors are areas adjacent to the first-order neighbors of area $i$; this logic continues for $k$th order neighbors. We assume the direct and indirect effects to be additive, so that $Z_i=z$ does not affect indirect effect estimation. We also note that this indirect effect is not comparable to the indirect effects estimated in mediation analysis. Kao termed these ``peer influence" causal effects\cite{Kao17}. In Figure~\ref{fig:ToyExample}, we can estimate the individual indirect effect by comparing the outcomes for area 3 between panels A and B or between panels C and D, where area 3's treatment is held constant in both these comparisons. We define the average total indirect effect of a unit's $k$th-order neighbors as:
\begin{equation}\label{eq:ie}
\overline{IE}=\frac{1}{N}\sum_{i=1}^N Y_i(Z_i=z,f(\mathbf{Z}_{\mathbb{N}_i}=\mathbf{z}))-Y_i(Z_i=z,f(\mathbf{Z}_{\mathbb{N}_i}=\mathbf{0})).
\end{equation}

If we consider an outcome $\mathbf{Y}$ that is the number of outcome events occurring in an area, e.g. number of deaths, number of cancer cases, etc., then we can model the individual potential outcomes for area $i$ as:
\begin{equation}\label{Poisson}
Y_i(Z_i,f(\mathbf{Z}_{\mathbb{N}_i}))\sim \mbox{Poisson}(\theta_iE_i)
\end{equation}
where $\theta_i$ is the relative risk of event $Y$ happening in unit $i$ and $E_i$ is the expected number of events in unit $i$. We further model the relative risk as:
\begin{equation}\label{eq:logtheta}
\log(\theta_i)=\tau\cdot Z_i+\gamma \cdot f(\mathbf{Z}_{\mathbb{N}_i})+\sum^P_{p=1}\beta_{p}{X_{ip}}+\epsilon_i
\end{equation}
where $\mathbf{X}$ represent $P$ confounders and $\epsilon\sim N(0,\sigma^2)$ is iid random effect. We note that $\tau=\overline{DE}$ and $\gamma=\overline{IE}$. We also assume no interaction between the direct and spillover effects.

\subsection{Ignorability Under Interference}
In RCM without interference, we typically assume $P(Z| Y(0),Y(1),X)=P(Z|X)$. In observational studies, it is often assumed that covariates $\mathbf{X}$ are sufficient to adjust for confounding in the treatment-outcome relationship. In other words, the potential outcomes are independent of the treatment assignment conditional on the covariates. With interference, we must assume that the potential outcomes $\mathbf{Y(z)}$ are independent of the treatment assignment  $\mathbf{Z}$ given the covariates $\mathbf{Z}$ {\em and} the influence network $\mathbf{A}$. That is,
\begin{equation}\label{A1}
\mathbf{Y(z)}\Perp \mathbf{Z}| \mathbf{X}, \mathbf{A} \tag{A1}.
\end{equation}
We make the additional assumption that the potential outcomes are independent of the influence network conditional on the covariates, or
\begin{equation}\label{A2}
\mathbf{Y(z)}\Perp \mathbf{A}| \mathbf{X} \tag{A2}.
\end{equation}
Kao called these assumptions the unconfounded treatment assignment and network influence assumptions under network interference\cite{Kao17}. We collectively call assumptions \ref{A1} and \ref{A2} ignorability under interference.

Practically, ignorability under interference entails defining both a treatment assignment mechanism and an interference mechanism. Covariates relevant to both mechanisms need to be identified. Characterizing a treatment assignment mechanism is well established in the literature\citep{ImbensGuido2015}, but the interference mechanism is a novel concept and should be viewed as ``the vehicle through which exposures to peer treatments are delivered" \citep{Kao17}. That is, the interference mechanism is how the treatment status of one unit affects the other units. The covariates necessary to meet unconfoundedness for both mechanisms will depend on the data and research question, and the relevant covariates may overlap. We partition the relevant covariates $\mathbf{X}$ as $[\mathbf{X}_Z,\mathbf{X}_A]$ to reflect this.

Rubin first outlined the use of Bayesian inference for estimating causal effects \cite{Rubin78}. Under the Bayesian framework, potential outcomes are treated as random variables and can be partitioned as $[\mathbf{Y}_{obs},\mathbf{Y}_{mis}]$ for the observed and missing potential outcomes. Bayesian imputation is used to compute the posterior distribution for the missing potential outcomes and causal estimands of interest. Ignorability under interference simplifies modeling and inference by dropping $\mathbf{Z}$ and $\mathbf{A}$ from the posterior distribution.

\subsection{Propensity Scores}

While ignorability under interference allows us to estimate causal effects with interference, we must still address other challenges to estimating causal effects that occur in observational studies such as covariate imbalances. Because treatment is not randomized in an observational study, we expect treated units to display different characteristics than the control units. This covariate imbalance must be addressed in order to estimate causal effects. Propensity scores are typically estimated as predicted values from a logistic regression model where the outcome is a binary variable indicating which units receive treatment $\mathbf{Z}$. The logistic regression model should adjust for the covariates $\mathbf{X}$ relevant to the treatment assignment and interference mechanisms. Treated and control units are likely to be similar if they have the same propensity score value. By controlling for covariates relevant both to the treatment assignment and to the interference mechanism, we expect units to be similar both in the probability of receiving treatment and in the probability of spillover exposure based on those covariate values. We illustrate this by assuming that the treatment assignment $\mathbf{Z}$ is independent of the potential outcomes given the covariates $[\mathbf{X_Z,X_A}]$ (\ref{A1}), or
\begin{center}
$P(Z=1|Y(Z),X_A,X_Z)=P(Z|X_A,X_Z)$,
\end{center}
and that the influence network $\mathbf{A}$ is independent of the potential outcomes given the covariates $\mathbf{X_A}$ (\ref{A2}), or
\begin{center}
$P(A|Y(Z),X_A)=P(A|X_A)$.
\end{center}
Then it follows that
\begin{center}
$P(Z=1| Y(Z),X_A,X_Z, PS(X_A,X_Z))= P(Z=1| PS(X_A,X_Z))$
\end{center}
where $PS(X_A,X_Z)$ is the propensity score that accounts for the covariates $[\mathbf{X_Z,X_A}]$. The proof extends from Imbens and Rubin\citep{ImbensGuido2015} and uses iterated expectations:
\begin{align*}
P(Z=1| Y(Z),X_A,X_Z, PS(X_A,X_Z))&=\mathbb{E}_Z(Z| Y(Z), X_A,X_Z, PS(X_A,X_Z))\\
&= \mathbb{E} [\mathbb{E}_Z\{Z|Y(Z), X_A,X_Z, PS(X_A,X_Z)\}| Y(Z), PS(X_A,X_Z)]\\
&= \mathop{\mathbb{E}} [\mathbb{E}_Z\{Z| X_A,X_Z, PS(X_A,X_Z)\}| Y(Z), PS(X_A,X_Z)]\\
&= \mathbb{E} [\mathbb{E}_Z\{Z| PS(X_A,X_Z)\}| Y(Z), PS(X_A,X_Z)]\\
&= \mathbb{E}_Z| PS(X_A,X_Z)\}\\
&=P(Z=1| PS(X_A,X_Z) ).
\end{align*}
In other words, given a vector of covariates that ensure unconfoundedness $[\mathbf{X_Z,X_A}]$, adjusting for the treatment difference in a propensity score model removes all biases associated with differences in the covariates because, conditional on the propensity score, the treatment assignment should be independent of the covariates. Rosenbaum and Rubin showed that the difference in outcomes for treated and control units at the same propensity score value is an unbiased average treatment effect\cite{Rosenbaum83}. Propensity score use to address confounding in spatial settings has been limited \citep{Papadogeorgou18}.  

Once propensity scores are estimated for a study population, we consider three methods to ensure balance between treated and control units: i.) pruning; ii.) grouping; and iii.) matching. Pruning involves omitting units where there is no propensity score overlap; in other words, the omitted units have no comparable features to the comparison group. Grouping, or stratifying, involves identifying subgroups of treated and control units with similar propensity score values after pruning \citep{Rosenbaum83,Hong06}. Matching occurs when units are matched $1:n$ where $n$ is the number of matched units to a particular unit \citep{Stuart10}, e.g. we may match one treated unit to two control units so the analysis dataset will include $3\times n$ units ($n$ treated units and $2\times n$ control units). In matching, both $n$ and a caliper need to be set. The caliper avoids matching dissimilar units beyond a specified threshold. We expect that if the propensity score model includes the covariates $\mathbf{X}$ necessary for ignorability under interference, then we may account for pertinent differences between treated and control units and improve the credibility of estimated causal effects. We compare propensity score methods to traditional outcome regression of the log relative risk. In outcome regression, we keep all units from the analysis dataset, and we control for $\mathbf{X}$ as confounders with linear adjustment terms. We argue that inference from data with limited overlap may extrapolate causal effects with no observed information for certain confounders for certain units. Bayesian inference may handle some of this uncertainty, but propensity score methods reduce biasedness in causal estimates. The tradeoff is that as our study population changes and we shrink our sample size to reduce covariate imbalance between units, causal estimates can only be interpreted within that study population.

\section{Simulation Studies}
\label{S:3}
\subsection{Data Generation}
\begin{figure}[h!]
 \includegraphics[width=\textwidth]{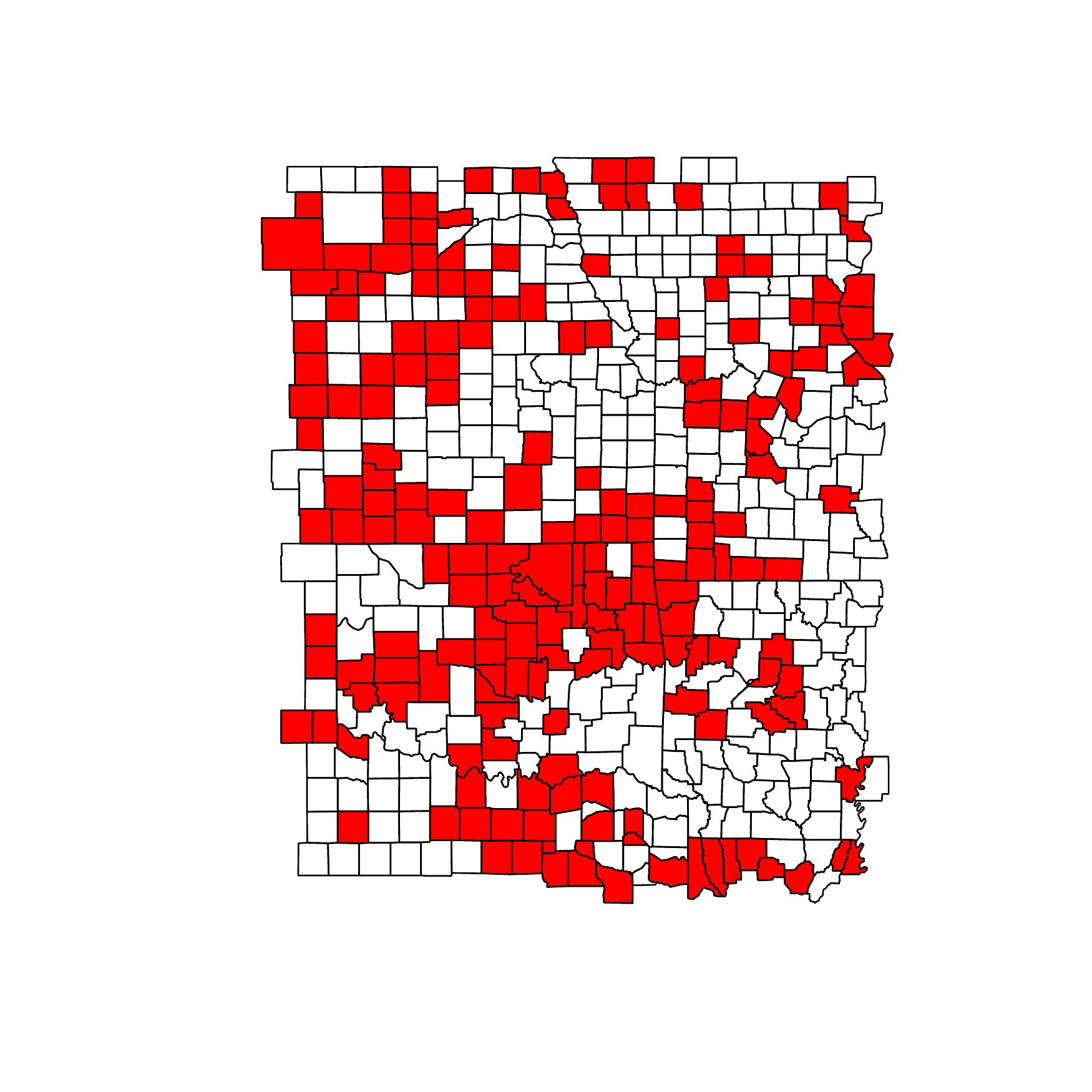}
\caption[Simulated study area.]{The study area of $N=500$ areal units where the red units indicate the treated units from one simulation.}
\label{fig:TreatedArea}
\end{figure}

We used a simulation study to compare the effectiveness of propensity score methods to outcome regression when estimating causal effects. We performed the simulation under a variety of scenarios that we describe below. In general, we simulated data for $N=500$ areal units based on real counties at the geographic center of the contiguous United States (Figure \ref{fig:TreatedArea}). We conducted 100 simulations for each scenario. For each of the 100 simulated datasets, we specified nine covariates $\mathbf{X}=X_1,\ldots,X_9$ for the 500 fixed units. For each covariate, we randomly generated a mixture distribution of treated and control units using package \textsf{distr} in \textsf{R} with overlap specified in Table~\ref{tab:covoverlap} \cite{R,distr}. We computed the empirical overlap between the treated and control distributions for each covariate by generating 1,000 values from each specified distribution using package \textsf{overlapping}. We report the average percentage overlap over the 1,000 generations in Table~\ref{tab:covoverlap}\cite{overlapping}. We also incorporated spatial autocorrelation into each covariate by computing a simultaneous autoregressive (SAR) weights matrix using function \textsf{invIrW} in library \textsf{spdep} and multiplying each generated variable by the matrix with a specified spatial dependence parameter (Table~\ref{tab:covoverlap}) \cite{Ord81,spdep}.

\begin{table}[h]
\caption[Simulated covariates for simulation.]{Simulated covariates where $r$ and $p$ are the failure and success probability rates for the negative binomial distribution respectively, $\alpha$ and $\beta$ are the shape and scale parameters for the gamma distribution respectively, $\mu$ and $\sigma$ are the mean and standard deviation parameters for the normal distribution respectively, and $\pi$ is the probability parameter for the Bernoulli distribution.}\label{tab:covoverlap}
\centering
\scalebox{0.9}{
\begin{tabular}{lp{37mm}lp{30mm}cp{15mm}c p{15mm}c}
\hline
\textbf{Covariate} & \textbf{\shortstack{Treated\\Distribution ($\mathbf{X_T}$)}} & \textbf{\shortstack{Control\\Distribution ($\mathbf{X_C}$)}} & \textbf{\shortstack{Spatial\\Autocorrelation}} & \textbf{\shortstack{Empirical\\Overlap (\%)}}\\
\hline
$X_1$ & NB($r=1.9, p=0.5$) & NB($r=2, p=0.4$) & 0.90 & 84.42 \\
$X_2$ & $\Gamma(\alpha=2,\beta=1$) & $\Gamma(\alpha=1,\beta=1$) & 0.70 & 79.66 \\
$X_3$ & N($\mu=5, \sigma=1$) & N($\mu=3,\sigma=2$) & 0.50 & 95.62 \\
$X_4$ & N($\mu=5, \sigma=2$) & N($\mu=3, \sigma$=1.5) & 0.60 & 55.29 \\
$X_5$ & $\Gamma(\alpha=8,\beta=1$) & $\Gamma(\alpha=1,\beta=3$) & 0.75 & 59.33 \\
$X_6$ & Bern($\pi=0.5$) & Bern($\pi=0.9$) & 0.30 & 74.17 \\
$X_7$ & N($\mu=10, \sigma=1.2$) & N($\mu=8, \sigma$=1.3) & 0.40 & 44.46 \\
$X_8$ & $\Gamma(\alpha=2, \beta=2$) & $\Gamma(\alpha=3, \beta=2$) & 0.50 & 77.80 \\
$X_9$ & $\Gamma(\alpha=2, \beta=2$) & N($\alpha=3, \beta=1$) & 0.60 & 48.32 \\
\hline
\end{tabular}}
\end{table}

Five of the simulated covariates, $X_1,\ldots,X_5$, served as confounders for the treatment assignment $\mathbf{Z}$. We represent the mixture distributions for the confounders as $\mathbf{X_{All}}$; the treated distributions for the confounders as $\mathbf{X_T}$; and the control distributions for the confounders as $\mathbf{X_C}$. We generated treatment $\mathbf{Z}$ as a binary variable where we modeled $\text{logit}(\Pr(Z_i=1))$. We also generated a population distribution for each unit as a negative binomial-distributed random variable with size 0.058 and mean 152,169, which we obtained from the applied dataset described in the next subsection.

\subsection{Specifying Treatment and Outcome}
In our simulations, we compared two treatment assignment models. First, we specified $\mathbf{Z}$ as a Bernoulli trial and modeled the logit of $\mathbf{Z}$ using the confounders' mixture distributions $\mathbf{X_{All}}$, which we modeled as
\begin{equation}\label{eq:trtmodel1}
\text{logit}(\Pr(Z_i=1))=\mathbf{X_{All,i}}\alpha+u_i+ -0.05 \cdot e_i
\end{equation}
where $\alpha=[0.0001, 0.00075, 0.005, -0.015,-0.001]$ and $u_i\sim \mbox{Normal}(0,1)$ and was multiplied by the SAR matrix with autocorrelation 0.9 to mimic unobserved spatial confounding in the treatment assignment and the random effect $e_i\overset{iid}\sim \mbox{Normal}(-0.5,1)$. We centered the error at -0.05 to allow the treatment assignment model to produce datasets with approximately 40.8\% of the 500 units receiving treatment in each simulation. As a second model, we specified $\mathbf{Z}$ using the covariate distributions of the treated units, $\mathbf{X_T}$, instead of $\mathbf{X_{All}}$, expressed 
\begin{equation}\label{eq:trtmodel2}
\text{logit}(\Pr(Z_i=1))=\mathbf{X_T}\alpha+u_i-0.5\cdot e_i.
\end{equation}
where all other values remained the same as in \ref{eq:trtmodel1}. Using only the distributions of the confounders of the treated units to specify $\mathbf{Z}$ reflects real-life observational studies.

As we assumed the outcome $\mathbf{Y}$ to be Poisson-distributed, we modeled the log relative risk of $\theta_i$ and assumed a population risk of 0.001 so that the expected count in each area was $E_i=0.001\times Population_i$. We specified the outcome with and without interference. Under no interference, we generated relative risks for each unit as
\begin{equation}\label{eq:outcome1}
\log(\theta_i)=\tau\cdot Z_i.
\end{equation}
Under interference, we included a spillover effect $\gamma$ and specified the relative risk model as
\begin{equation}\label{eq:outcome2}
\log(\theta_i)=\tau\cdot Z_i+\gamma\cdot f(\mathbf{Z}_{-i})
\end{equation}
where  $f(\mathbf{Z}_{-i})$ was the proportion of neighbors receiving treatment $\mathbf{Z}$. In all instances, we specified the true direct causal effect as $\tau=3$ and the true spillover effect as $\gamma=2$ in simulations with interference. 

\subsection{Estimation}
The aim of our simulation study was to evaluate the impact of several factors on estimating direct and spillover effects. In our simulations that had no interference, i.e. we modeled the log relative risk of $\mathbf{Y}$ without interference (equation \ref{eq:outcome1}), we aimed to only estimate the direct effect, $\hat{\tau}$. In simulations with interference, we modeled the log relative risk of $\mathbf{Y}$ as in equation \ref{eq:outcome2} and estimated both $\hat{\tau}$ and the spillover effect, $\hat{\gamma}$, except in cases where we sought to assess the effect of ignoring interference. We compared estimation across four methods. The first of these methods was standard outcome regression as previously described. The remaining models were the propensity score-based methods: pruning, stratification, and matching. For the propensity score-based methods, we consider two different models: (1) the propensity score model (PSM) and (2) the potential outcomes model, which we model as $\log{\hat{\theta_i}}$. No PSM was used for standard outcome regression.

In order to achieve ignorability under interference, the relevant covariates for the treatment assignment mechanism consisted of the confounders $\mathbf{X}_Z=X_1,\ldots,X_5$. The relevant information for the interference mechanism, $\mathbf{X}_A$, was simply first-order proximity. We included this information explicitly when we modeled the spillover effect by only estimating the spillover effect in first-order neighboring units of other treated units. This is true for how we specified interference in the true outcome (equation \ref{eq:outcome2}).

For the outcome regression method (which we label ``Full" in the results, Tables~\ref{tab:probcov} and ~\ref{tab:mse}), we modeled the potential outcomes for the entire sample ($N=500$) in a regression model that assumed a linear relationship between the covariates and log-linear relative risk of the outcome. For simulations that did not consider interference, we modeled the log relative risk as
\begin{equation}\label{eq:reg1}
\log(\hat{\theta}_i)=\hat{\tau}\cdot Z_i+\sum^P_{p=1}\beta_pX_{ip}.
\end{equation}
Alternatively, for simulations that aimed to capture interference, we modeled the log relative risk as
\begin{equation}\label{eq:reg2}
\log(\hat{\theta}_i)=\hat{\tau}\cdot Z_i+\hat{\gamma}\cdot f(\mathbf{Z}_i)+\sum^P_{p=1}\beta_pX_{ip}.
\end{equation}
We determined which covariates $\mathbf{X}_p$ to adjust for in the outcome regression model by assessing the balance for a particular covariate between the treated and control unit using an unequal variance Student's $t$-test. We always adjusted based on the covariates' mixture distribution because in real life we cannot readily separate the treated and control units' covariate distributions.

For the pruning method (labeled ``Pruned" in Tables~\ref{tab:probcov} and ~\ref{tab:mse}) , we estimated the propensity score for each unit via logistic regression as
\begin{equation}
\text{logit}(\Pr(Z_i=1))=\mathbf{X} \hat{\alpha}
\end{equation}
where the $P$ covariates contained in $\mathbf{X}$ varied depending on the simulation setup (described in next section). We iteratively dropped units where there was no propensity score overlap and re-fit a PSM until all units were within the propensity score overlap. We used these $N_{pruned}$ units to model $\log(\hat{\theta}_i)$ where we adjusted only for the covariates that were not balanced within the $N_{pruned}$ units using the mixture distributions.

For the stratified propensity score method (labeled ``Stratified" in Tables~\ref{tab:probcov} and ~\ref{tab:mse}), we used the same $N_{pruned}$ units and categorized the treated and control units into quintiles based on their propensity score. We adjusted for the strata in the potential outcomes model along with any covariates that were not balanced within the strata, i.e. we modeled
\begin{equation}
\log(\hat{\theta}_i)=\hat{\tau}\cdot Z_i+\hat{\gamma}\cdot f(\mathbf{Z}_i)+\sum^P_{p=1}\hat{\beta}_pX_{ip}+\sum^5_{d=1}\hat{\upsilon}_d,
\end{equation}
where $\hat{\upsilon}_d$ is the linear adjustment for each propensity score strata. To determine covariate selection, we calculated Student's $t$-test for each covariate in each strata and evaluated whether the $|\max(t)|>1.96$ where $t$ is the observed test statistic; if so, we included that particular covariate in the potential outcomes model. We chose quintiles due to frequent use in the literature \citep{Rosenbaum83,Austin11}.

In the matching method (labeled ``Matched" in Tables~\ref{tab:probcov} and ~\ref{tab:mse}), we matched treated and control units 1:1 with a caliper of 1.0 standard deviations of the propensity score. That is, we matched only units where the difference in propensity score values did not exceed more than one standard deviation of the PSM. Because not every unit can be matched this way, the sample size was reduced to $N_{matched}$.

We estimated all models using Markov chain Monte Carlo (MCMC) in \textsf{OpenBUGS} and the \textsf{R} computing environment \cite{doi:10.1002/sim.3680}. For each model, we ran two chains for 20,000 iterations with a burn-in of 10,000 and thinned the remaining sample to every 15 iterations. We provided non-informative normal priors centered at 0 with variances equal to 1,000 for each parameter such that
\begin{equation}
\hat{\tau},\hat{\gamma},\hat{\beta}_p, \hat{\upsilon}_d \sim \mbox{Normal}(0,1000).
\end{equation}
We assessed the chains' convergence using the Gelman-Rubin Diagnostic \cite{gelman1992}. If the chains did not converge, then we iteratively ran the chains for another 20,000 iterations with the same burn-in and thinning parameter until convergence was achieved. We assessed the mean posterior estimate for the direct and spillover effects and the variance of the distribution in order to calculate the mean squared error (MSE) and probability coverage for the estimates relative to the known true values. We considered the optimal method to be the one with highest probability coverage and minimal MSE.

\subsection{Simulation Scenarios}
Our simulations explored a variety of possible scenarios when estimating causal effects with and without interference (Table~\ref{tab:simsetup}). In scenarios 1 and 3a, we explored how interference affects estimation of the direct and spillover effects by modeling the log relative risk of outcome $\mathbf{Y}$ without interference using equation \ref{eq:outcome1}. For all other scenarios, we incorporated interference by modeling $\mathbf{Y}$ as specified in equation \ref{eq:outcome2}. We further explored the impact of different treatment assignment models in scenarios 1, 2a, and 2b compared to all other scenarios. In scenarios 1, 2a, and 2b, we generated treatment assignment $\mathbf{Z}$ as in equation \ref{eq:trtmodel1} where the confounders' mixture distribution informed the treatment assignment generation. In the other scenarios, only the confounder distributions of the treated units contributed to $\mathbf{Z}$, as specified in equation \ref{eq:trtmodel2}. In all cases, the PSM relies on the confounders' mixture distributions as in practice.

\begin{table}[h]
\caption[Simulation scenarios outline.]{Outline of scenarios simulated. Each column represents factors that were altered between scenarios.}
\label{tab:simsetup}
\centering
\scalebox{0.75}{
\begin{tabular}{ccccccc} 
\hline
\multicolumn{1}{c}{Scenario} & \multicolumn{1}{c}{\begin{tabular}[c]{@{}c@{}}Treatment\\Assignment\\ Covariates\end{tabular}} & \multicolumn{1}{c}{\begin{tabular}[c]{@{}c@{}}Interference\\ Present?\end{tabular}} & \multicolumn{1}{l}{Estimands} &
 \multicolumn{1}{c}{\begin{tabular}[c]{@{}c@{}}Spatial Random\\ Effect Estimated?\end{tabular}} & \multicolumn{1}{c}{\begin{tabular}[c]{@{}c@{}}Additional\\Covariates\\ in PSM?\end{tabular}} & \multicolumn{1}{c}{\begin{tabular}[c]{@{}c@{}}Missing\\Confounders\\ in PSM?\end{tabular}} \\ \hline
1       & $\mathbf{X}_{All}$ &  No     &   $\hat{\tau}$    & No  &    -     &  -      \\
2a      & $\mathbf{X}_{All}$ &  Yes     &  $\hat{\tau}$   &  No &   -     &    -    \\
2b      & $\mathbf{X}_{All}$ &  Yes     &  $\hat{\tau},\hat{\gamma}$   & No  &    -    &     -   \\
3a      & $\mathbf{X}_T$  & No     &  $\hat{\tau}$  &   No &  -      &   -     \\
3b      & $\mathbf{X}_T$  &  Yes     &  $\hat{\tau},\hat{\gamma}$     & No  &    -    &   -     \\
4a     &$\mathbf{X}_T$   &  Yes     &  $\hat{\tau}$      & \textbf{Yes}   &    -    &   -     \\
4b    & $\mathbf{X}_T$  &  Yes     &  $\hat{\tau}, \hat{\gamma}$     & \textbf{Yes}   &     -   &    -    \\
5      & $\mathbf{X}_T$   &  Yes     &  $\hat{\tau}, \hat{\gamma}$    &   No & $X_6,\ldots,X_9$        &    -    \\
6a      & $\mathbf{X}_T$   &  Yes     &  $\hat{\tau},\hat{\gamma}$     & No   &   -    & Omit $X_5$       \\
6b      & $\mathbf{X}_T$  &  Yes     &  $\hat{\tau}, \hat{\gamma}$   &  No &    -   & Omit $X_4$ and $X_5$                                                                             \\
\hline
\end{tabular}}
\end{table}

From this paradigm, scenario 1 may be described as a straightforward causal analysis with no interference, considered as a baseline case. Scenario 3a was a similar straightforward causal analysis with no interference, but illustrated the effects of the treatment assignment $\mathbf{Z}$ being specified as it truly is in nature, i.e. treatment assignment is informed only by the confounders' treated distributions versus the observed mixtures.

Scenarios 2a and 2b described estimating the direct effect when there is interference and the analysis ignores it (2a) and when the analysis assumes or tackles it (2b). In both these scenarios, we assumed $\mathbf{Z}$ was generated using the confounders' mixture distribution. Scenario 3b was arguably the most straightforward and realistic scenario to practice where there was interference, the analysis assumed it, and the treatment assignment was informed by only the confounders' treated distributions. Scenario 3a contrasted from 3b by having no interference and the analysis not assuming it.

We evaluated the use of spatial random effects in scenarios 4a and 4b by including such an effect in the potential outcomes model. We were interested in the random effects' ability to discern a spillover effect when we did and did not specify a spillover effect explicitly in our potential outcomes model (equation \ref{eq:reg1} versus \ref{eq:reg2}, or 4a versus 4b). In both scenarios, we had specified the true outcome with interference. We specified the spatial random effect to have an intrinsic conditionally autoregressive prior (CAR) \citep{Besag1991}, which conditions a variable on its neighbors' values using a spatial adjacency matrix. The CAR random effect $\mathbf{\zeta}$ may be specified as:
\begin{equation}
\zeta_i| \zeta_{-i}, \mathbf{W}, \omega^2 \sim N \Bigg(\frac{\sum^N_{i=1}w_{ji}\zeta_i}{\sum^N_{i=1}w_{ji}}, \frac{\omega^2}{\sum^N_{i=1} w_{ji}} \Bigg)
\end{equation}
where $\mathbf{W}$ is a $N \times N$ adjacency matrix with $w_{ij}=1$ if areas $i$ and $j$ are spatially contiguous and $\omega^2$ is the variance of the spatial random effects $\mathbf{\zeta}$.

In scenario 5, we considered the addition of covariates beyond the true confounders $\mathbf{X}_T$. Specifically, we introduced  $X_6,\ldots,X_9$ to the outcome regression model of the log relative risk and to the PSMs for each method. In contrast, we considered the omission of true confounders from the analysis in scenarios 6a and 6b. In 6a, we omitted $X_5$ from the outcome regression and PSM, and in 6b, we omitted both $X_4$ and $X_5$.

\subsection{Results}
As expected, we found that the average sample size shrunk as we moved from the full dataset to the pruned to the matched dataset in our simulation study. For the full method, which used the entire simulated sample data, the analysis dataset was consistently the 500 counties. However, we found the pruned and stratified datasets to be around 495 units as shown in Table~\ref{tab:nsim}. The matched datasets showed the most drastic size reduction. The lowest sample sizes observed were in scenarios 6a and 6b ($\overline{N}_{Matched}=391.50$) and scenario 5 ($\overline{N}_{Matched}=391.90$). These scenarios modified the number of covariates in the PSM, which in turn may have increased the difficulty of matching by reducing the propensity score overlap between treated and control units by either using less covariates (scenarios 6a and 6b) or incorporating irrelevant covariates (scenario 5). This contributed to more dissimilar treated and control units.

\begin{table}[h]
\caption[Sample size for each scenario.]{The average sample size for each method from the 100 simulations for each scenario.}\label{tab:nsim}
\resizebox{\textwidth}{!}{ 
\begin{tabular}{ l r r r r r r r r r r }
\hline
\textbf{Sample Size} & \textbf{1} & \textbf{2a} & \textbf{2b} & \textbf{3a} & \textbf{3b} & \textbf{4a} & \textbf{4b} & \textbf{5} & \textbf{6a} & \textbf{6b}\\
\hline
$\overline{N}_{Pruned}$ & 495.43 & 495.43 & 495.43 & 495.72 & 495.72 & 495.75 & 495.75 & 494.41 & 494.59 & 495.93\\
$\overline{N}_{Matched}$ & 416.50 & 416.50 & 416.50 & 411.72 & 411.72 & 410.49 & 410.40 & 391.90 & 391.50 & 391.50 \\
\hline
\end{tabular}}
\end{table}

We found that the stratified and matched methods consistently outperformed the full and pruned methods when estimating direct and spillover effects simultaneously. In scenario 1, which contained no interference and only attempted to estimate a direct effect, we found the higher probability coverages in the pruned and matched methods (91\% and 90\% respectively, Table~\ref{tab:probcov}); however, the matched method had the lowest MSE (1.012 $\cdot10\times E^{-5}$, Table~\ref{tab:mse}). That is, the credible interval for the direct effect $\hat{\tau}$ contained the true value of $\tau$ for 90 of the 100 simulations using the matched dataset. In comparison, the credible interval for the direct effect only contained the true value for 65 of the 100 simulations using the full dataset.

In scenario 2a, we introduced interference in the dataset, but made no attempt to estimate a spillover effect. In turn, we observed that we entirely missed estimating the true direct effect (0\% probability coverage across all methods), though the MSE remained relatively low (Table~\ref{tab:mse}). In scenario 2b, we estimated both the direct and spillover effects when interference was present. The full method had marginal probability coverage when estimating the direct effect (26\%), but found most success in the stratified and matched methods. The credible interval for the direct effect contained the true value of $\tau$ 94 times out of the 100 simulations for both methods (Table~\ref{tab:probcov}). Similarly, the credible interval for the spillover effect had 91\% and 92\% probability coverage for the stratified and matched methods, respectively.

In scenarios 3a onward, we found that covariate imbalance played a role in successful estimation of the direct and spillover effects. The full method continued to underperform the other methods even when there was no interference, with lower probability coverage for capturing the true direct effect and higher MSE (scenario 3a in Tables~\ref{tab:probcov} and \ref{tab:mse}). When interference was introduced, the full method severely failed: the credible interval for $\hat{\tau}$ only captured the true direct effect 33 of the 100 simulations in scenario 3b. The full method was slightly better at estimating the spillover effect; the credible intervals for $\hat{\gamma}$ contained the true spillover effect for 52 of the 100 simulations.

Spatial random effects, as demonstrated in scenarios 4a and 4b, did not improve estimation for the direct and spillover effects. When interference was present and a spillover effect was not included in the potential outcomes model (scenario 4a), the probability of capturing the true direct effect was 50\% or lower for all methods (Table~\ref{tab:probcov}) and the MSE was five-fold higher when compared to all other scenarios (Table~\ref{tab:mse}). For the pruning scenario, the MSEs may be highest because we dropped observations but still needed to estimate coefficients for imbalanced covariates. Even when the spillover effect was included in the potential outcomes model (scenario 4b), we compared scenario 4b to 3b to find that the model with a spatial random effect did not perform as well as the model with no spatial random effect. In fact, the matched method with a spatial random effect captured the true direct effect 69 out of 100 times compared to the matched method with no spatial random effect, which captured the true direct effect 91 out of 100 times. This was also true when estimating the spillover effect: 64\% probability coverage compared to 94\% probability coverage.

Incorporating extraneous covariates into the PSM increased the probability coverage for the direct effect when estimating the direct and spillover effects in the full method (scenario 5 compared to scenario 3b). The stratified method had the highest probability coverage for estimating both effects, however (Table~\ref{tab:probcov}). We observed slightly higher MSE for estimating these effects when the PSM contained irrelevant confounders (Table~\ref{tab:mse}).

The MSE increased even more in scenarios 6a and 6b as we omitted relevant confounders ($X_5$ in scenario 6a and both $X_4$ and $X_5$ in scenario 6b). However, we still found high probability coverage (above 84\%) for capturing the true direct and spillover effects in every method except the full method in these scenarios.

\begin{table}[h]
\caption[Probability coverage for each scenario.]{The probability coverage for each simulation scenario. Each probability corresponds to the number of times out of 100 simulations that the method resulted in a 95\% credible interval for the direct effect $\tau$ and spillover effect $\gamma$, when appropriate, that contained the true known value. We bolded the highest probability coverage corresponding to each method for each parameter.}\label{tab:probcov}
\centering
\scalebox{0.9}{
\begin{tabular}{ l l r r r r r r r r r r }
\hline
\textbf{Parameter} & \textbf{Method} & \textbf{1} & \textbf{2a} & \textbf{2b} & \textbf{3a} & \textbf{3b} & \textbf{4a} & \textbf{4b} & \textbf{5} & \textbf{6a} & \textbf{6b}\\
\hline
Direct & Full & 65.0 & 0 & 26.0 & 70.0 & 33.0 & 5.0 & 49.0 & 55.0 & 44.0 & 40.0 \\
 & Pruned & {\bf 91.0} & 0 & 87.0 & 91.0 & 83.0 & 28.0 & 53.0 & 91.0 & 84.0 & 88.0 \\
 & Stratified & 83.0 & 0 & {\bf 94.0} & 85.0 & {\bf 94.0} & {\bf 50.0} & {\bf 75.0} & {\bf 95.0} & 94.0 & 85.0 \\
 & Matched & 90.0 & 0 & {\bf 94.0} & \bf{92.0} & 91.0 & 1.0 & 69.0 & 87.0 & {\bf 96.0} & {\bf 93.0} \\
Spillover & Full & - & - & 51.0 & - & 52.0 & - & 45.0 & 71.0 & 61.0 & 67.0 \\
 & Pruned & - &- & \bf{93.0} & - & 93.0 & - & {\bf 68.0} & 91.0 & 86.0 & 91.0 \\
 & Stratified  & - & - & 91.0 & - & {\bf 96.0} & - & 37.0 & {\bf 94.0} & 89.0 & {\bf 96.0} \\
 & Matched  & - & - & 92.0 & - & 94.0 & - & 64.0 & 88.0 & {\bf 93.0} & 91.0 \\
\hline
\end{tabular}}
\end{table}

\begin{table}[h]
\caption[Mean squared error for each scenario.]{The mean squared error (MSE) corresponding to estimates for the direct effect and spillover effect, when appropriate, for each simulation scenario. We report MSE scaled by $10\times E^{-5}$ except where \textsuperscript{$\Psi$}, which we report as unscaled. We bolded the lowest MSE corresponding to each method for each parameter.}\label{tab:mse}.
\centering
\scalebox{0.8}{
\begin{tabular}{ l l r r r r r r r r r r }
\hline
\textbf{Parameter} & \textbf{Method}  & \textbf{1} & \textbf{2a} & \textbf{2b} & \textbf{3a} & \textbf{3b} & \textbf{4a}\textsuperscript{$\Psi$} & \textbf{4b}\textsuperscript{$\Psi$} & \textbf{5} & \textbf{6a} & \textbf{6b} \\
\hline
Direct & Full & 8.386 & {\bf 0.408} & 32.689 & 6.951 & 32.727 & 4.821 & 7.016 & 10.974 & 22.835 & 18.495 \\
 & Pruned & 1.671 & 1.214 & 5.309 & 1.233 & 6.692 & 17.893 & 56.888 & 2.899 & 5.586 & 2.965 \\
 & Stratified & 1.874 & 0.649 & {\bf 1.595} & 1.762 & {\bf 1.658} & {\bf 2.009} & {\bf 1.244} & 9.174 & 3.326 & 2.231 \\
 & Matched & {\bf 1.012} & 1.406 & 4.876 & {\bf 0.358} & 2.168 & 3.117 & 5.767 & {\bf 2.519} & {\bf 1.440} & {\bf 1.859} \\
Spillover & Full & - & - & 7.526 & - & 7.245 & - & 3.594 & 3.880 & 6.169 & 5.645 \\
 & Pruned& - & - & 1.382 & - & 2.105 & - & 56.888 & 1.396 & 1.953 & 1.232 \\
 & Stratified  & - & - & 1.322 & - & 1.123 & - & {\bf 3.388} & {\bf 1.076} & 1.656 & 1.381 \\
 & Matched  & - & - & {\bf 1.258} & - & {\bf 0.985} & - & 8.532 & 1.518 & {\bf 1.022} & {\bf 0.999} \\
\hline
\end{tabular}}

\end{table}

\section{Data Application}
\label{S:4}
\subsection{Methods}
We applied the methods to publicly available data from the Surveillance, Epidemiology, and End Results (SEER) program and Environmental Protection Agency (EPA) to estimate the causal effects of air pollution regulation on lung cancer incidence. SEER is a cancer registry database covering approximately 26\% of the U.S. population with data registries including California, New Mexico, Iowa, Kentucky, Louisiana, Connecticut, New Jersey, and Georgia along with several metropolitan areas. The database contains information on year and county of cancer diagnosis and patient demographics\citep{Cahoon15}.

In 1970, the Clean Air Act allowed the EPA to regulate air emissions and establish National Ambient Air Quality Standards (NAAQS) intended to protect public health and manage hazardous pollutant emissions\cite{EPA16}. In 1990, under the Clean Air Act Ammendments (CAAA), the EPA began to regulate air quality emissions and designate areas in violation of the NAAQS as ``nonattainment", prompting these areas to take actions to improve air quality\citep{Zigler12}. In 1997 the EPA updated its standards for ambient concentrations of particulate matter with aerodynamic diameter $<2.5 \mu m$ (PM$_{2.5}$). In 2005, the EPA designated certain counties as nonattainment or otherwise ``attainment" (or ``unclassifiable" if there was not enough data to classify\citep{EPA16b}). The EPA made these designations based on nine factors including historic air quality, population density and urbanization, traffic and commuting patterns, meterology, and geography \cite{EPACh5}. They used available data from 2001 onward.

\begin{figure}[h]
 \includegraphics[width=\textwidth]{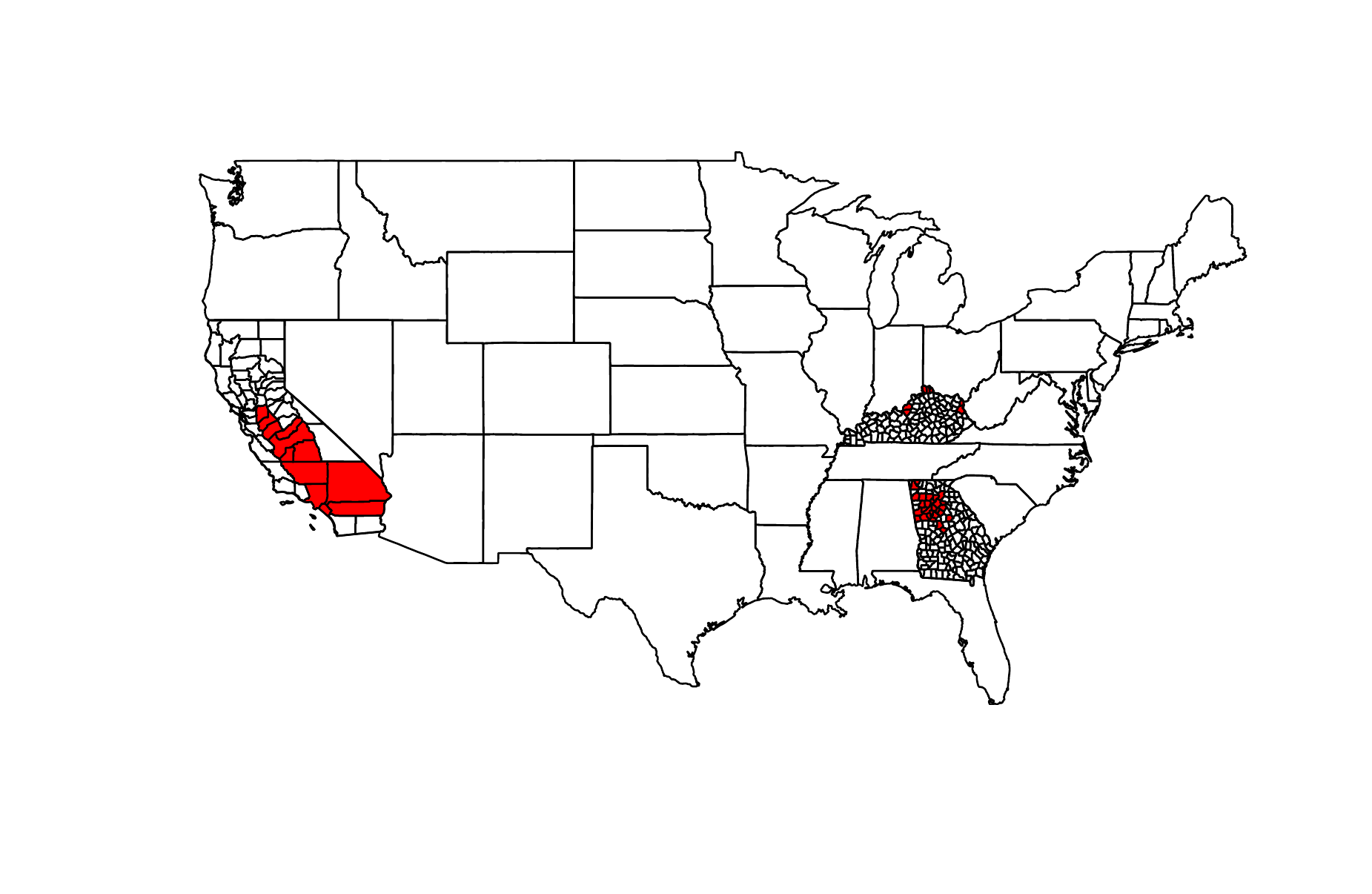}
\caption[SEER Study Area.]{The study area of $N=337$ counties in California, Georgia, and Kentucky where counties designated as nonattainment by the U.S. Environmental Protection Agency for 1997 National Ambient Air Quality Standards for particulate matter with aerodynamic diameter $<2.5 \mu$ m (PM$_{2.5}$) are shaded red.}
\label{fig:StudyArea}
\end{figure}

Several of these regulated areas overlap with the study area of the SEER Program. Our analysis focused specifically on lung cancer cases in California, Georgia, and Kentucky (Figure~\ref{fig:StudyArea}) based on the ratio of nonattainment counties to all counties respective to each state. Multiple studies have established significant associations between air pollution exposure and lung cancer incidence \cite{GharibvandLida2017TAbA, VilleneuvePaulJ.2014VeaR,HartJaimeE.2014ICES}. We evaluated lung cancer cases reported between 2005 and 2013. We aimed to estimate the causal effects of the 1997 (PM$_{2.5}$) nonattainment designations on lung cancer incidence starting in 2005 following nonattainment designation. Based on previous research on particulate matter with aerodynamic diameter $<10$ $\mu m$ and ozone (O$_3$)\citep{Zigler12}, we expected interference and sought to estimate direct causal effects in the nonattainment counties and spillover causal effects in surrounding counties.

We imputed PM$_{2.5}$ values from the Downscaler Model\cite{Downscaler}. The Downscaler Model allowed for finer scale predictions rather than simply observed values from the EPA Air Quality System, because the Downscaler Model utilizes both observed measurements and the Community Multi-Scale Air Quality Model (CMAQ). Because the data are reported as a mean value with a standard error at the census tract level, we treated PM$_{2.5}$ as a random Gaussian process with mean and standard error specified by the Downscaler Model. For each ZIP code, we generated ten values in each year and aggregated to the county level using the mean. The final analysis was performed using the county-level aggregated data.

We obtained additional data from multiple sources including nonattainment designations from the EPA Nonattainment Areas for Criteria Pollutants (``Green Book")\cite{EPA16b}, population demographics from the U.S. Census, meteorological variables from the Automated Surface Observing System (ASOS), elevation at county centroids from ESRI, and county-level smoking rates estimated from the Centers for Disease Control and Prevention's Behavioral Risk Factor Surveillance System \cite{Dwyer-Lindgren2014}. We considered lung cancer cases reported between 2002 and 2004 and ASOS climate metrics observed prior to 2005 to be baseline measurements.

We obtained ASOS weather data from 705 weather stations located across the three states in our study area in addition to the bordering states for edge correction. The stations collected data with varying frequency (typically more than once a day) and with no consistent coverage across the study area. For each year, we averaged observed measurements for wind speed (knots), wind direction (degrees), relative humidity, dew point and air temperatures (Fahrenheit), visibility (in miles), barometric pressure, and hourly precipitation (inches). We omitted observations where wind exceeded 50 knots, relative humidity was observed over 100\%, and visibility was observed greater than 11 miles, believing such observations to be errors. To average circular wind direction, we converted wind speed and direction into U and V cosine direction vectors. For each variable, we interpolated observations at weather station sites across each state individually (California, Georgia, and Kentucky) using inverse distance weighting (IDW). We chose the optimal power and neighborhood size by comparing sums of squared residuals after 5-fold cross-validation. We also compared IDW to ordinary kriging and found that IDW had lower sum of squared residuals. We assigned each county the interpolated value at the county's centroid. Finally, we reverted the U and V vectors to wind direction and speed (which we report in meters per second). This interpolation method was shown effective for directional data by Gumiaux \cite{GumiauxC2003Gatb}.

The full dataset contained information on the 337 counties in the three states, of which 46 were designated nonattainment. To build the PSM for each method, we used Student's $t$-test for covariate selection and also considered practical knowledge of nonattainment designation and PM$_{2.5}$ dispersion. While we made no inference from the PSM, we limited the number of covariates in the model to avoid overfitting and used mean values across the study period, i.e. we averaged values between 2002 and 2004 for pre-treatment periods and between 2005 and 2013 for post-treatment. The final PSM contained the covariates listed in Table~\ref{tab:confounders}. Of these, we identified confounders relevant to the treatment assignment to be a county's ozone nonattainment designations made in 2004 based on 8-hour levels; mean PM$_{2.5}$ between 2002 and 2004 and between 2005 and 2013, separately; mean temperature; percent urban housing units; mean relative humidity; elevation of the country centroid; population density in 2000; and the mean amount of time spent traveling to work by workers aged 16 or older (minutes). We also included dewpoint temperature in 2000 and 2013 as possible confounders based on imbalances found in exploratory analyses. For the interference mechanism, we identified mean wind speed between 2005 and 2013 and elevation as relevant covariates that could contribute to spillover. We chose these covariates to meet the ignorability under interference assumption. We believe the treatment (PM2.5 nonattainment designation) and the network influence mechanism to be unconfounded based on subject matter expertise, including the variables that determined nonattainment designations and the factors that influence downwind spillover.

In the full model, we used data from all 337 counties and modeled the log relative risk of lung cancer while adjusting for ozone nonattainment designation, mean PM$_{2.5}$, mean temperature, mean relative humidity, population density, percentage urban housing, mean work travel time, dewpoint temperature in 2000 and 2013, smoking prevalence, and the mean number of lung cancer cases 2002-2004.

To prune the data, we fit a logistic regression PSM that contained all confounders listed in Table~\ref{tab:confounders}. Within the pruned dataset, we checked for imbalances in the covariates before modeling the log relative risk in a potential outcomes model. We used the pruned dataset additionally for the stratified potential outcomes model. We identified three subgroups based on the PSM, splitting the data at the 86th and 96.5th quantiles of the PSM. We chose these quantiles in order to have an equal number of nonattainment counties within each strata. The potential outcomes model contained two indicators to correspond to these three subgroups. We again checked for covariate balance within these strata and adjusted for the appropriate confounders. Finally, we created two matched datasets from the pruned dataset. In the first matched dataset, we matched 1:1 with a caliper of 0.25 standard deviation of the PSM. In the second dataset, we matched 1:1 with a caliper of one standard deviation. Once more, we controlled for relevant confounders in the potential outcomes model for both matched datasets.

For all of the potential outcomes models, we included a direct effect term, $\hat{\tau}$, corresponding to nonattainment designation and a spillover term, $\hat{\gamma}$, corresponding to the proportion of nonattainment counties surrounding a county. We also included a temporal term for each year with an exchangeable prior to capture changes in lung cancer risk across time not captured by confounders or nonattainment designation. We provided normal priors to all estimated parameters with gamma hyperpriors for the variance terms with shape and scale parameters of 0.1 and 0.1, respectively. For each model, we ran two MCMC chains for the same lengths, burn-in, and thinning as the simulation studies until we achieved convergence. Overall, nonattainment designation was static based on the 2005 designations. For the final report models, we also assumed the spillover structure to be the proportion of adjacent nonattainment counties bordering a county.

We considered several spillover structures for this application. We initially assumed spillover would be directional based on wind patterns. We used the interpolated annual wind bearing for each county to identify first- and second-order neighboring counties of nonattainment counties whose centroids fell within 30$^\circ$ of the wind bearing. We expanded this threshold to 45$^\circ$ to compare whether this better captured spillover counties. We also evaluated two simpler spillover structures: an indicator for whether a county bordered a nonattainment county and the proportion of nonattainment counties bordering a county. We compared models' deviance information criteria (DIC)\cite{doi:10.1111/1467-9868.00353} to assess which spillover structure best fit the matched data. Models with lower DIC are considered better candidates for a dataset. In our evaluation, we also included models with spillover terms corresponding to a mixture of structures. We found that the proportion of nonattainment counties bordering a county as a spillover structure resulted in a model with the lowest DIC. We also found the weight for this structure to be near 1 in models that included this structure in weighted mixtures.

\begin{table}[h]
\caption[Descriptive statistics for covariates.]{Descriptive statistics for confounders that we deemed necessary for ignorability under interference. We report mean (standard deviation) for all continuous covariates except ozone nonattainment designation (*), which we report as the proportion of counties designated nonattainment for 8-hour ozone in 2004. For continuous covariates, we report the p-value from a two-sample Student's $t$-test. For ozone nonattainment designation, we report a p-value from a two proportion $z$-test.}\label{tab:confounders}
\centering
\scalebox{0.9}{
\begin{tabular}{ l l l l }
\hline
\textbf{Covariate} & \textbf{Nonattainment Counties} & \textbf{Control Counties} & \textbf{P-value} \\
\hline
Mean PM$_{2.5}$ 2002-2004 ($\mu$g/$m^3) $& 11.4 (2.18) & 11.4 (2.04) & 0.995   \\
Mean PM$_{2.5}$ 2005-2013  & 12.8 (0.69) & 11.1 (1.48) & $<$ 0.001   \\
Mean Work Travel Time (Minutes) & 28.3 (4.52) & 25.86 (5.02)  & 0.001   \\
Elevation (feet) & 327.3 (387.40) & 282.9 (369.49) & 0.470   \\
Smoking Prevalence (\%)  & 25.5 (4.04) & 28.7 (4.73) & $<$ 0.001 \\
Population Density 2000  & 580.6 (769.58) & 166.6 (993.63) & 0.002  \\
Mean Lung Cancer Cases 2002-2004  & 277.7 (619.60) & 50.0 (125.97)  & 0.017  \\
Mean Wind Speed 2005-2013  (meters/second) & 0.8 (0.55) & 0.7 (0.54) & 0.237   \\
Mean Temperature 2005-2013 ($^\circ$F)  & 60.88 (3.16) & 59.70 (5.01)  & 0.03   \\
Mean Relative Humidity (\%) & 69.0 (7.18) & 72.1 (4.01) & 0.006   \\
Dewpoint Temperature (2000) ($^\circ$F)  & 46.4 (3.65) & 48.3 (4.75) & 0.002  \\
Dewpoint Temperature (2013) ($^o$F)  & 46.3 (5.31) & 48.6 (6.38) & 0.011 \\
Migration Rate (Persons) & 7.4 (2.18) & 7.0 (3.79) & 0.301   \\
Urban Housing (\%)  & 76.5 (23.45) & 34.3 (29.24) & $<$ 0.001  \\
Ozone Nonattainment Designation* (\%) & 19.6 & 80.4 & $<$ 0.001 \\
\hline
\end{tabular}}
\end{table}

\subsection{Results}
\begin{figure}[H] 
  \begin{subfigure}{6cm}
    \centering\includegraphics[width=6cm]{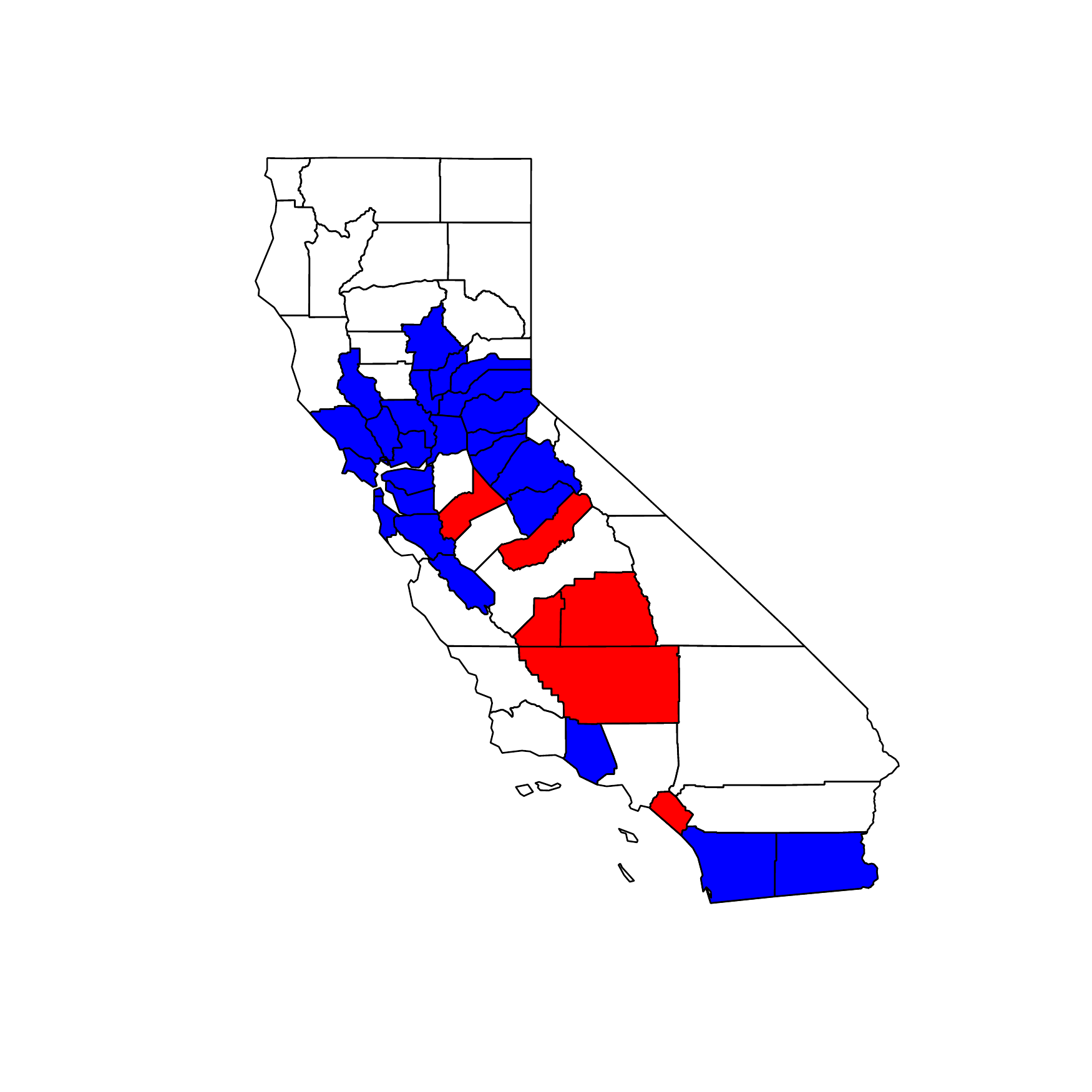}
    \caption{California}
  \end{subfigure}
  \begin{subfigure}{6cm}
    \centering\includegraphics[width=6cm]{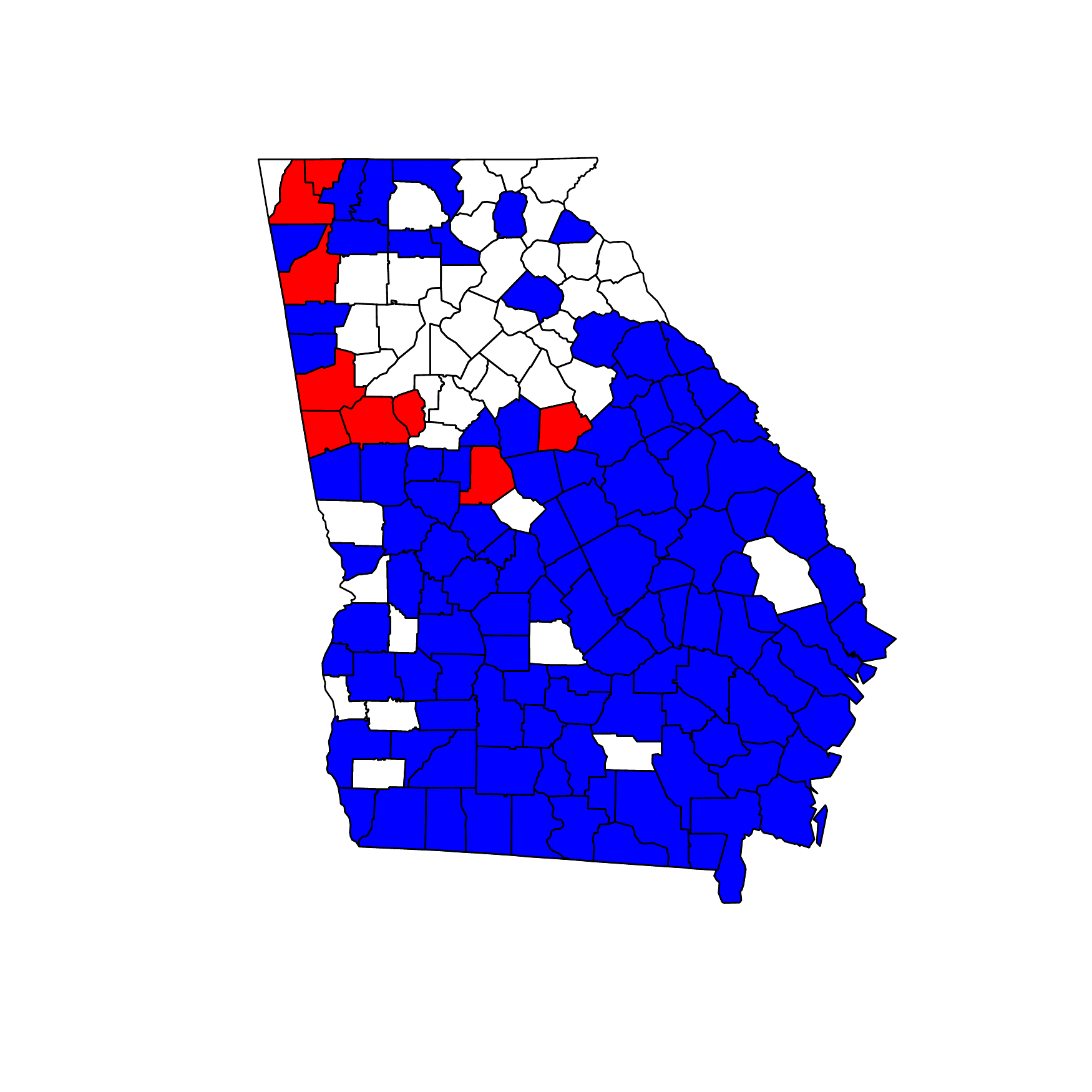}
    \caption{Georgia}
  \end{subfigure}

  \begin{subfigure}{6cm}
    \centering\includegraphics[width=6cm]{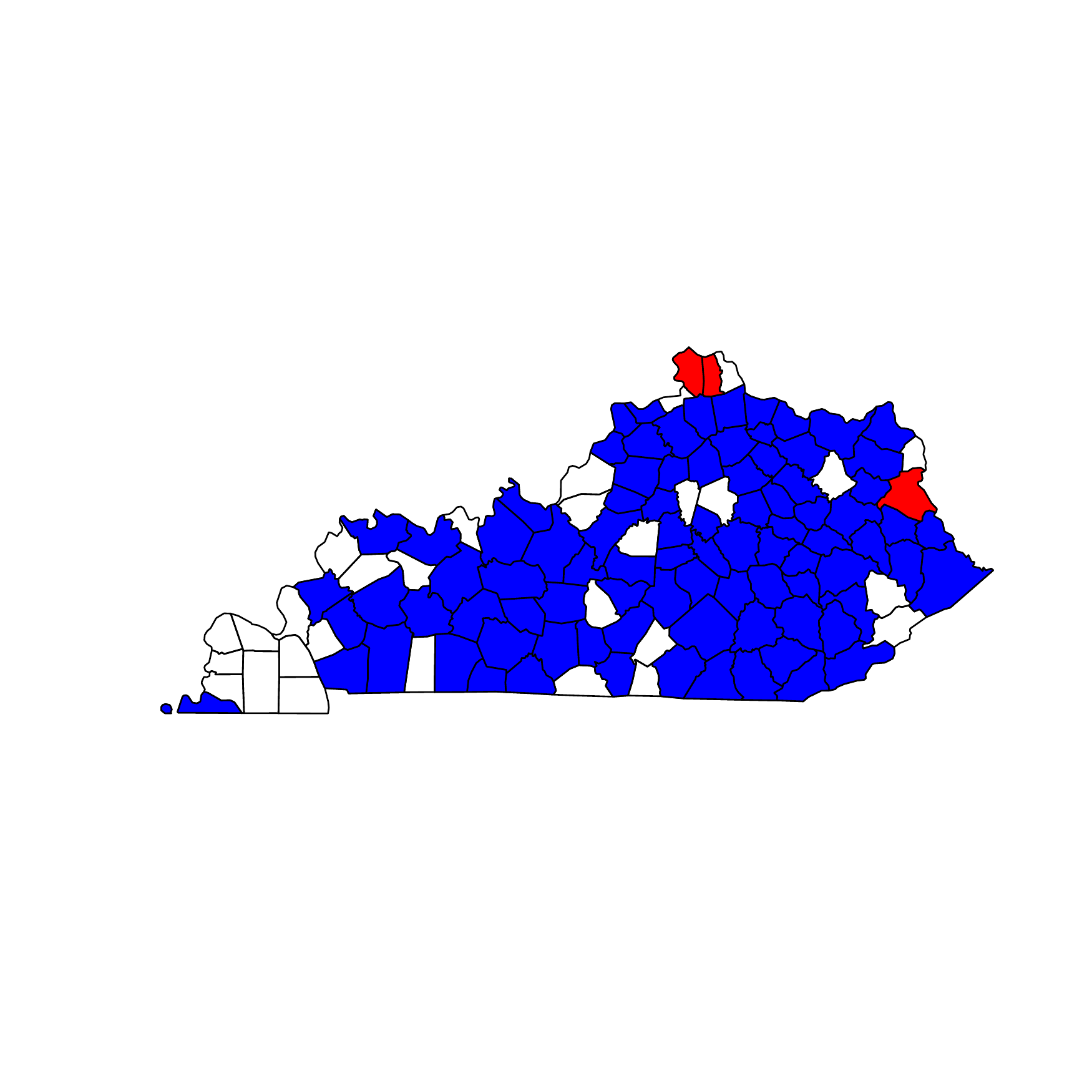}
    \caption{Kentucky}
  \end{subfigure}
\caption[Map of pruned data.]{The pruned dataset contains $N=239$ counties from California (a), Georgia (b), and Kentucky (c) (outlined in black). Counties colored red are retained in the pruned dataset and considered nonattainment while blue counties are retained and considered control.}\label{fig:pruneddata}
\end{figure}

Similar to the simulation study results, the sample size for the applied dataset shrank between methods. The pruned dataset contained 239 counties (Figure~\ref{fig:pruneddata}). We adjusted for ozone nonattainment designation, mean relative humidity, mean work travel time, elevation, and smoking prevalence in the pruned model along with a temporal term. For the stratified model, we used the same 239 counties and adjusted for mean relative humidity, dewpoint temperature in 2013, smoking prevalence, and ozone nonattainment designation in the potential outcomes model.

\begin{figure}[H] 
  \begin{subfigure}{6cm}
    \centering\includegraphics[width=6cm]{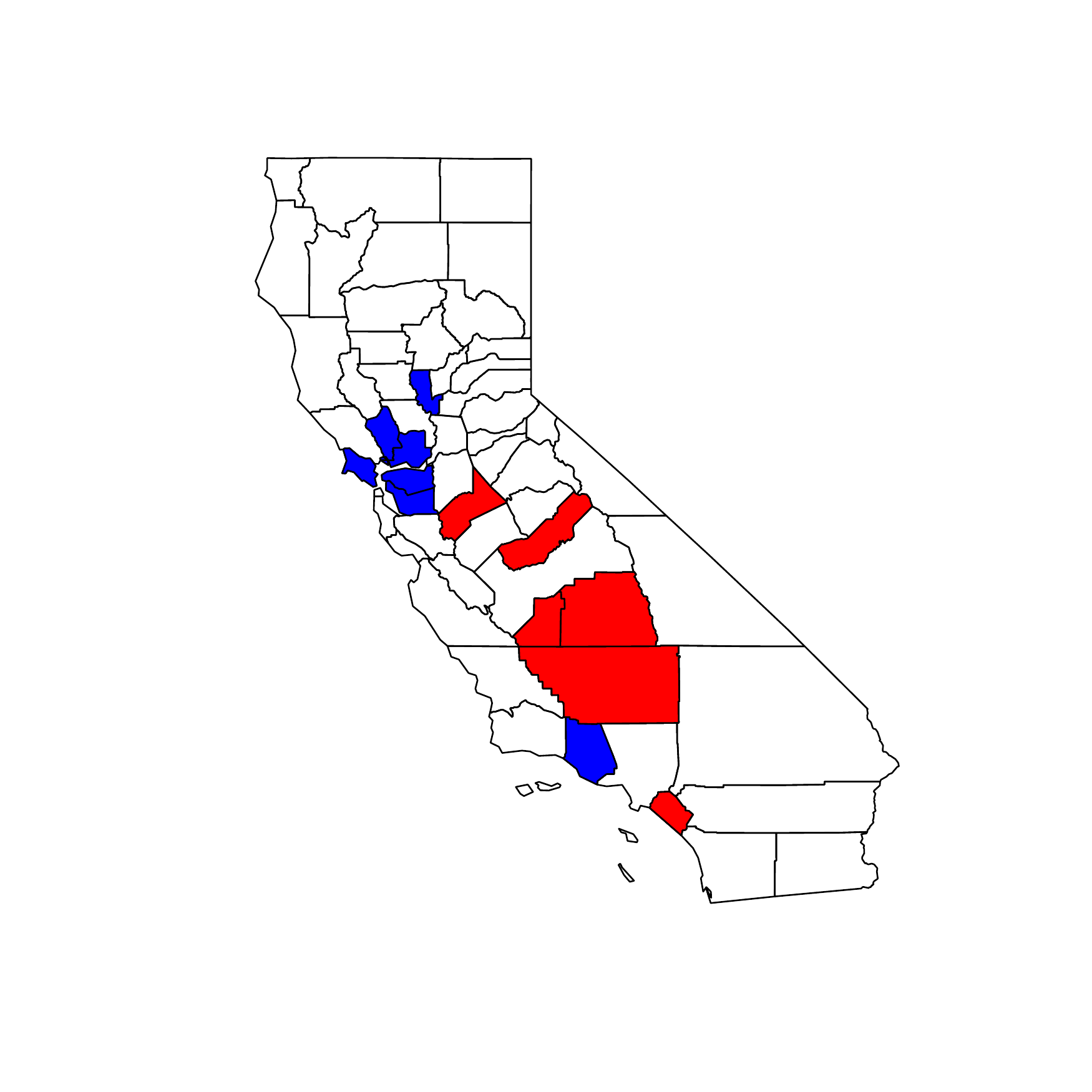}
    \caption{California}
  \end{subfigure}
  \begin{subfigure}{6cm}
    \centering\includegraphics[width=6cm]{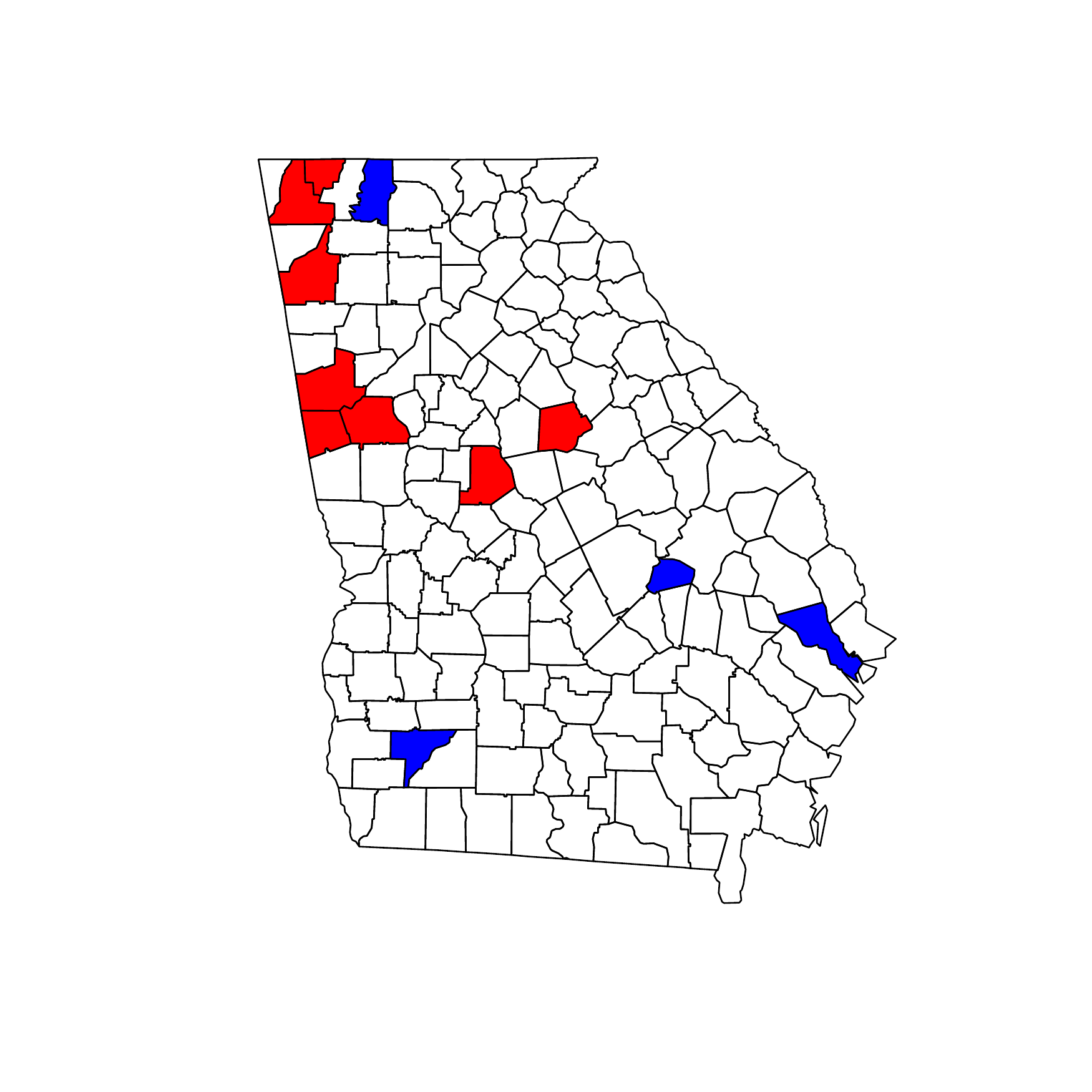}
    \caption{Georgia}
  \end{subfigure}

  \begin{subfigure}{6cm}
    \centering\includegraphics[width=6cm]{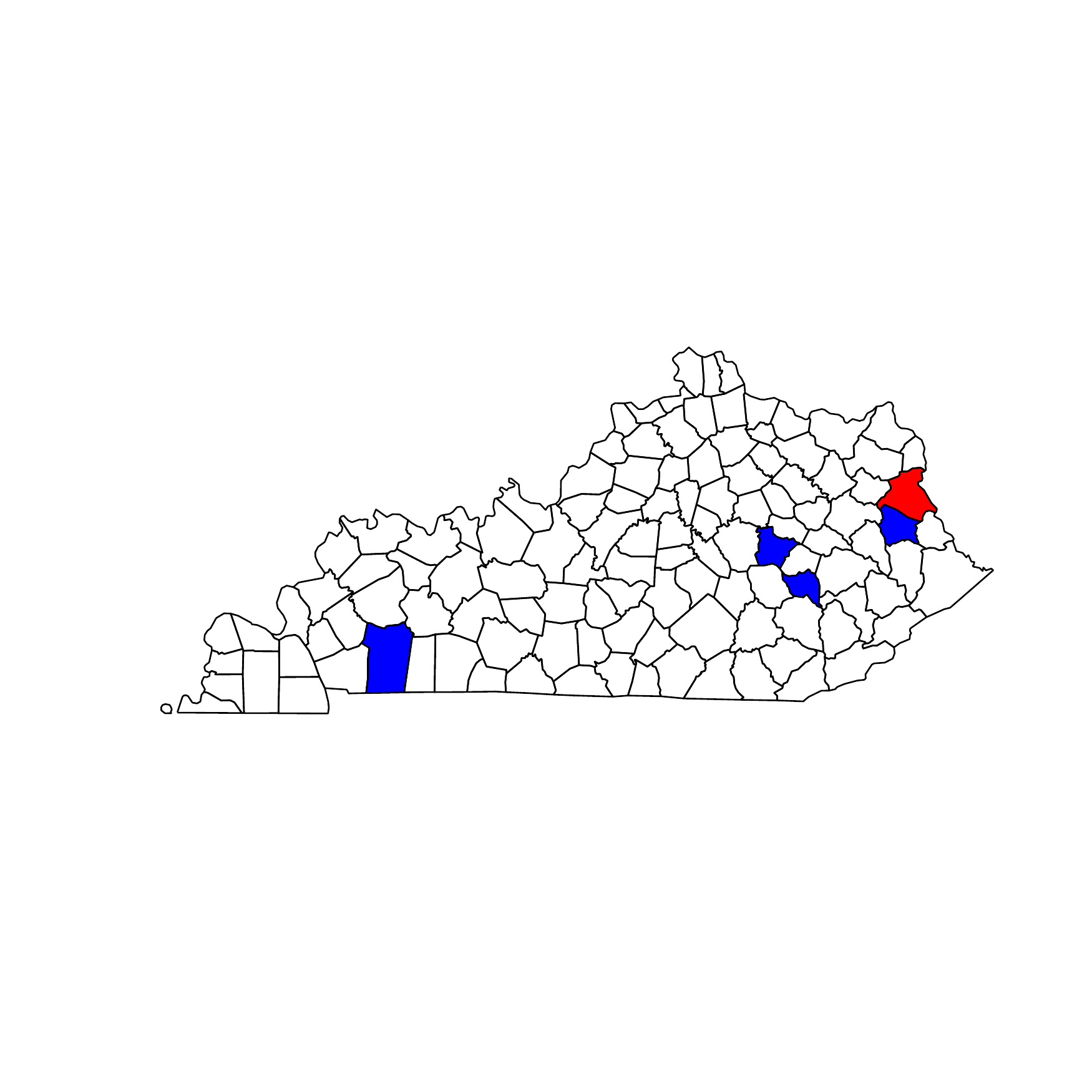}
    \caption{Kentucky}
  \end{subfigure}
\caption[Map of matched data ($N=30$).]{The matched dataset contains $N=30$ counties from California (a), Georgia (b), and Kentucky (c) (outlined in black). Counties colored red are retained in the pruned dataset and considered nonattainment while blue counties are retained and considered control.}\label{fig:matcheddata30}
\end{figure}

\begin{figure}[H]
  \begin{subfigure}{6cm}
    \centering\includegraphics[width=6cm]{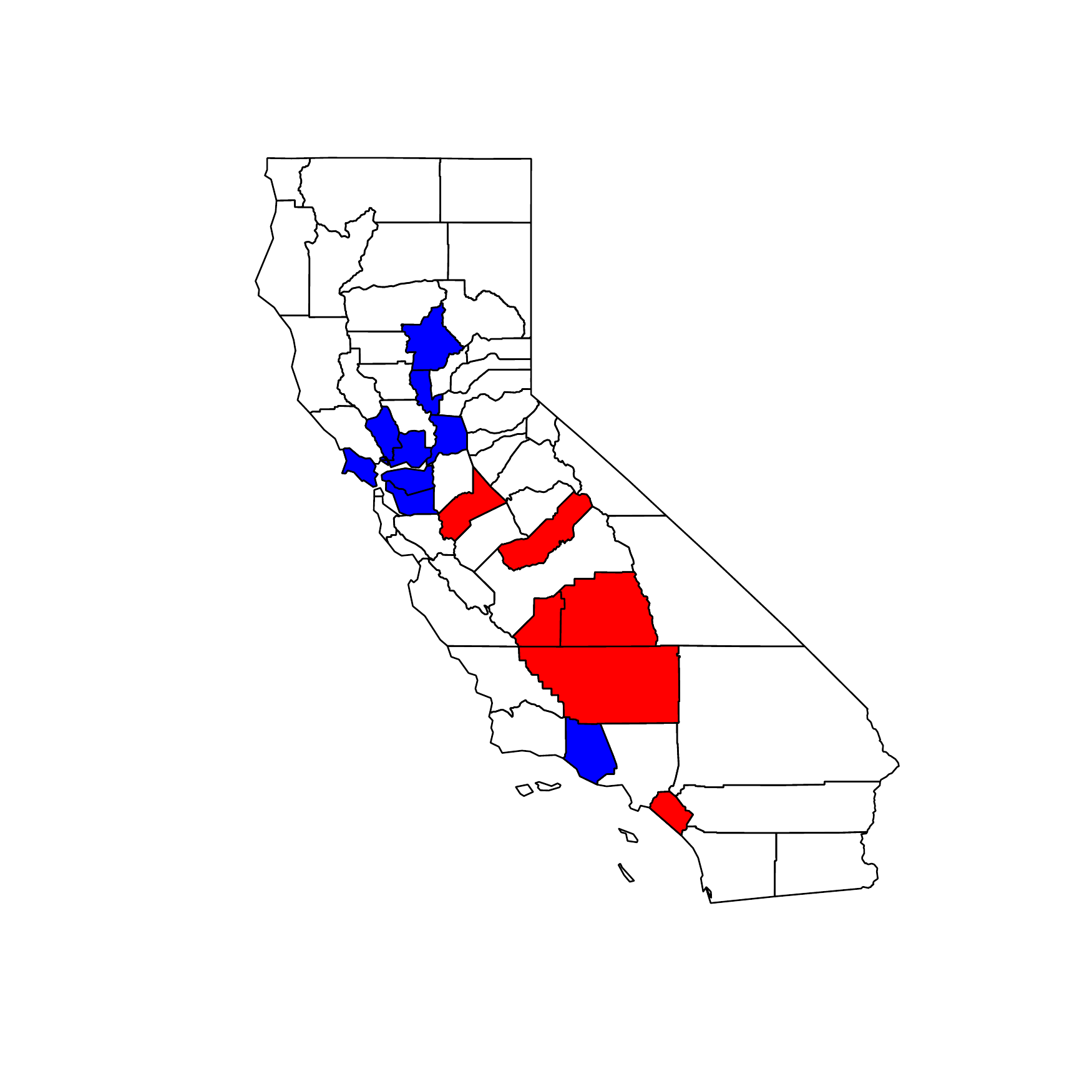}
    \caption{California}
  \end{subfigure}
  \begin{subfigure}{6cm}
    \centering\includegraphics[width=6cm]{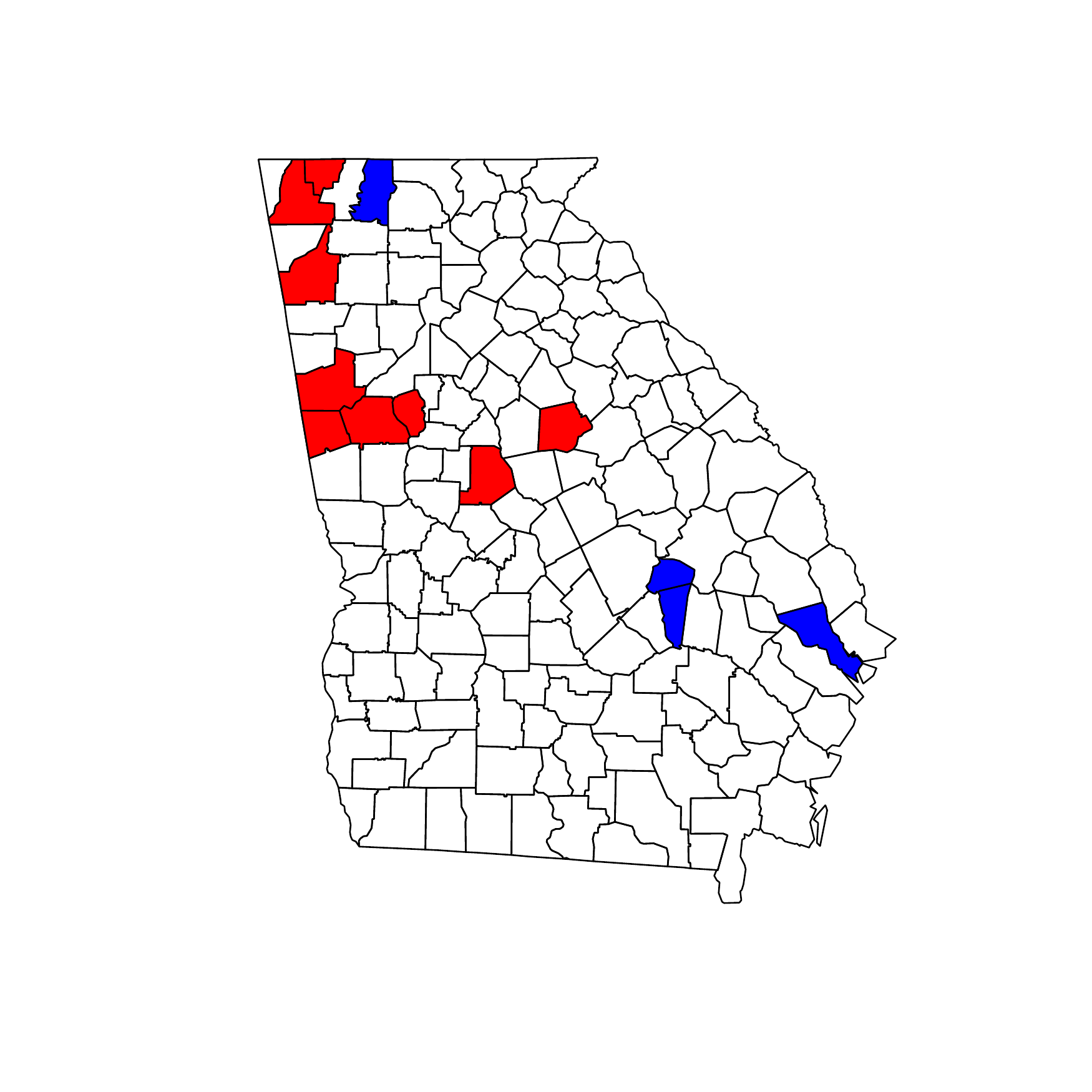}
    \caption{Georgia}
  \end{subfigure}

  \begin{subfigure}{6cm}
    \centering\includegraphics[width=6cm]{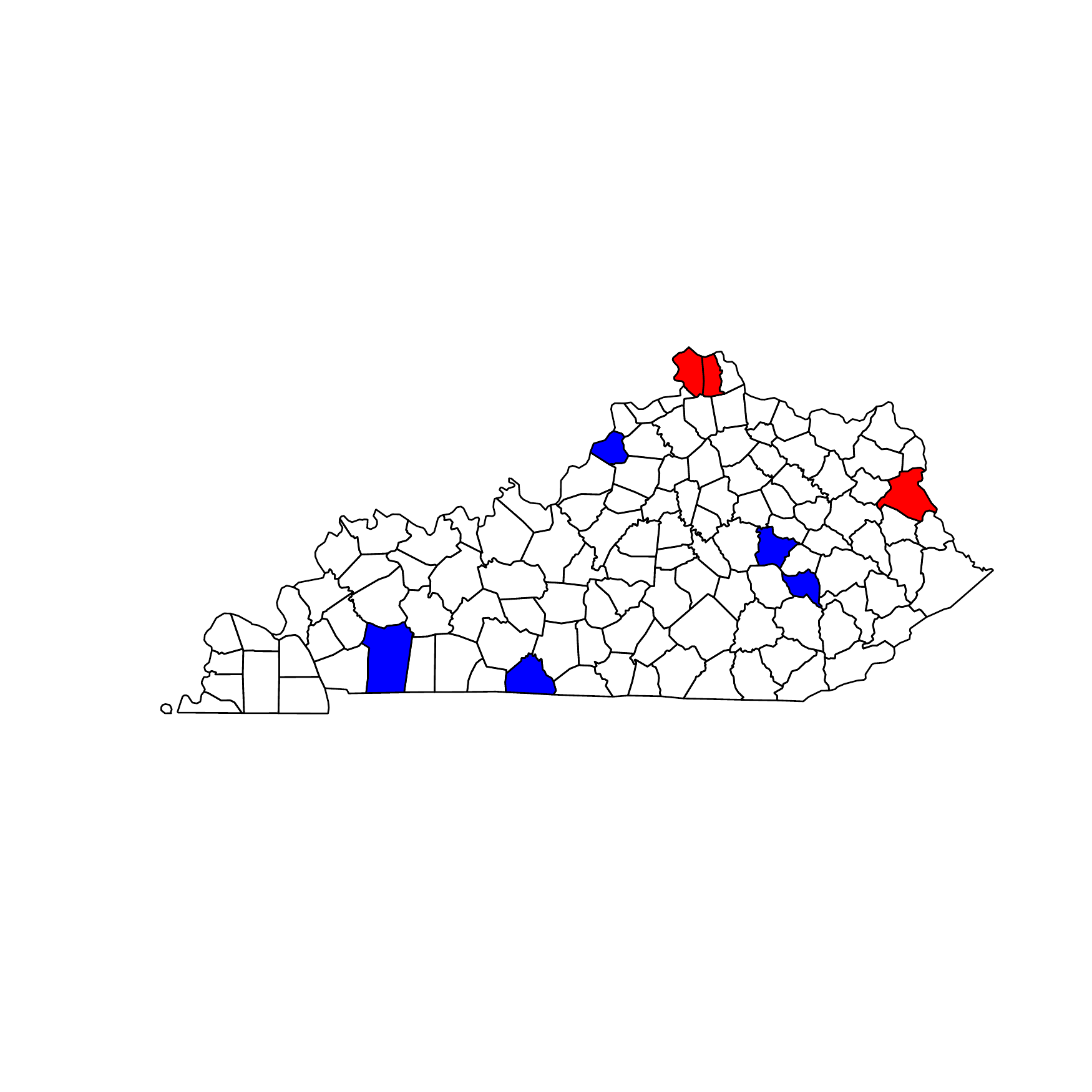}
    \caption{Kentucky}
  \end{subfigure}
\caption[Map of matched data ($N=36$).]{The matched dataset contains $N=36$ counties from California (a), Georgia (b), and Kentucky (c) (outlined in black). Counties colored red are retained in the pruned dataset and considered nonattainment while blue counties are retained and considered control.}\label{fig:matcheddata36}
\end{figure}

In the two matched models, we continued to control for mean PM$_{2.5}$ 2002-2004, smoking prevalence, and ozone nonattainment designation. The first matched dataset (caliper equal to 0.25 standard deviation) contained 30 counties (Figure~\ref{fig:matcheddata30}). The second matched dataset contained 36 counties due to the larger caliper (equal to 1 standard deviation) (Figure~\ref{fig:matcheddata36}).

In general, we found a direct and spillover effect that trended toward protective across the models (near or below 1 on the relative risk scale, Table~\ref{tab:AppResults}). We found a significant direct effect in the full and stratified models. According to the full model, counties with nonattainment designation have between 0.90 and 0.91 times the risk of lung cancer incidence as control counties after adjusting for relevant covariates and the proportion of nonattainment counties bordering a county. Similarly, the stratified model identified counties with nonattainment designation as having between 0.93 and 0.96 times the risk of lung cancer incidence compared to counties without nonattainment designation.

For the spillover effect, the risk trended toward protective except in the case of the full model (posterior mean: 1.66, 95\% CI: 1.61, 1.71). These results suggest a decreased risk of lung cancer incidence in counties bordering counties with nonattainment designation. This decreased relative risk would only be fully achieved if a county is entirely surrounded by nonattainment counties (of which only five counties in our entire dataset are); otherwise, this risk must be multiplied by the proportion of a county's neighbors that are designated nonattainment. The total causal effect is the sum of the direct and spillover effects. We remind the reader that the true effects could be not protective and misleading if we have not addressed all confounders, i.e., violated our ignorability under interference assumption. More research needs to occur to validate these findings. Further research may also consider including second- or higher-order neighbor spillover effects. However, due to counties' larger size compared to oft-used ZIP codes or census tracts or block groups in spatial analyses, first-order neighbor counties may capture substantial portions of PM$_{2.5}$'s range.

Because the matched sample sizes were so small, we conducted an additional simulation study to assess how sample size may affect causal effect estimation. Using the template of scenario 3b, which we believed to best approximate the applied analysis, we sampled 30 units from the complete matched dataset in each of the 100 simulations. We found probability coverages of 88\% and 93\% and MSE of $9.93\times 10^{-5}$ and 0.00019, respectively, for the direct and spillover effects. Based upon the results of this simulation study, the observed results in our application are credible.

\begin{table}[h]
\caption[Results from application to SEER data.]{Results from application to SEER data. We report posterior mean estimates (95\% credible interval) on the relative risk scale.}\label{tab:AppResults}
\centering
\resizebox{\textwidth}{!}{ 
\begin{tabular}{cccccc}
\textbf{Estimate}         & \textbf{Full Model}                                         & \textbf{Pruned Model}                                       & \textbf{Stratified Model}                                   & \textbf{\begin{tabular}[c]{@{}c@{}}Matched Model\\ (Caliper 0.25)\end{tabular}} & \textbf{\begin{tabular}[c]{@{}c@{}}Matched Model\\ (Caliper 1)\end{tabular}} \\
\hline
\textbf{N}                & 337                                                         & 239                                                         & 239                                                         & 30                                                                              & 36                                                                           \\
& & & & & \\
\textbf{Direct Effect}    & \begin{tabular}[c]{@{}c@{}}0.91\\ (0.90, 0.91)\end{tabular} & \begin{tabular}[c]{@{}c@{}}1.02\\ (1.00, 1.04)\end{tabular} & \begin{tabular}[c]{@{}c@{}}0.95\\ (0.93, 0.96)\end{tabular} & \begin{tabular}[c]{@{}c@{}}0.99\\ (0.97, 1.02)\end{tabular}                     & \begin{tabular}[c]{@{}c@{}}0.99\\ (0.97, 1.01)\end{tabular}                  \\
& & & & & \\
\textbf{Spillover Effect} & \begin{tabular}[c]{@{}c@{}}1.66\\ (1.61, 1.71)\end{tabular} & \begin{tabular}[c]{@{}c@{}}0.77\\ (0.75, 0.79)\end{tabular} & \begin{tabular}[c]{@{}c@{}}0.98\\ (0.95, 1.01)\end{tabular} & \begin{tabular}[c]{@{}c@{}}0.75\\ (0.72, 0.79)\end{tabular}                     & \begin{tabular}[c]{@{}c@{}}0.69\\ (0.66, 0.72)\end{tabular}\\
 \hline
\end{tabular} }
\end{table}

\section{Discussion}
\label{S:6}

In this chapter, we have adapted new assumptions to spatial settings so that we may estimate causal effects in the presence of spatial interference. We have illustrated how these assumptions may be met when addressing causal questions in air pollution epidemiology and shown how to apply these assumptions to a specific dataset. We also proposed methods to accurately and precisely estimate direct and spillover causal effects in the presence of interference and have demonstrated the scenarios in which these methods may outperform each other.

In general, we argue that researchers should characterize both a treatment assignment and interference mechanism when dealing with data under interference \cite{Kao17}. We also encourage using propensity scores to handle imbalances in confounders, which are inherent in observational studies \cite{Stuart10,Austin11}.
The use of propensity scores in spatial causal analysis has been limited. We know of only one other paper that utilizes propensity scores to address spatial confounding\cite{Papadogeorgou18}, and in that case distance between study units was utilized within the propensity score model.

We have also shown that including spatial random effects in a potential outcomes model is inadequate for addressing spatial confounding. Even when the spillover is explicitly and correctly modeled, we found higher bias and variance in the estimates for direct and spillover effects if the analysis even captured the true parameter value at all. Spatial random effects have been used previously to address spatial interference\cite{Zigler17}, but our simulations show this may lead to misestimation of direct and spillover effects. Estimating multiple spatially structured terms, including a treatment effect, spillover effect, and spatial random effect, may be too computationally challenging as shown in scenarios 4a and 4b. Similar results were previously demonstrated by Hodges and Reich when studying associations \cite{HodgesReich10}, but were not explored in a causal context. They found that including spatially-correlated error terms may contribute to collinearity and inflate error variances. Further research may explore how including both a spillover effect corresponding to an asymmetric, directional neighborhood matrix and a CAR random effect based on first-order adjacency may interact together. We suspect, however, that CAR random effects are not so flexible that they could capture the structure of the indirect effect as we specified in our simulations (i.e., the proportion of treated neighbors). We also did not address in this paper how well CAR random effects estimate spatial dependence in data that was generated using a SAR, though using a binary weights matrix as we did is more similar to a CAR structure than row-standardized weights.

We also found that incorporating covariates that are not necessarily confounders, as may happen when the treatment assignment mechanism is not fully understood, does not severely hinder estimation of causal effects (scenario 5). Future may research may include a simulation where true confounders are included in the outcome model; we omitted this detail for simplicity's sake, yet illustrated how confounders may be modeled for estimation purposes. We also demonstrated that successful estimation is still possible when known confounders are omitted as long as covariate imbalances are appropriately addressed (scenarios 6a and 6b). In summary, rigorous methods may supplement missing confounders that theoretically violate ignorability under interference.

Further research may extend our simulation designs to covariates that are spatially correlated with one another. This is the first simulation design we know of to incorporate spatially autocorrelated covariates when assessing methods, but we also know that in reality covariates are often correlated with one another. This paper does not address how that may affect analysis. We also do not investigate how the size of an areal unit may affect results, though covariates such as population density and land mass may proxy for this information. We recognize that the size of an area, specifically the proportion of area bordering neighboring areas, may impact spillover effects and the accuracy of spillover effect estimation.

\begin{figure}[h!]
  \begin{subfigure}{6cm}
    \centering\includegraphics[width=6cm]{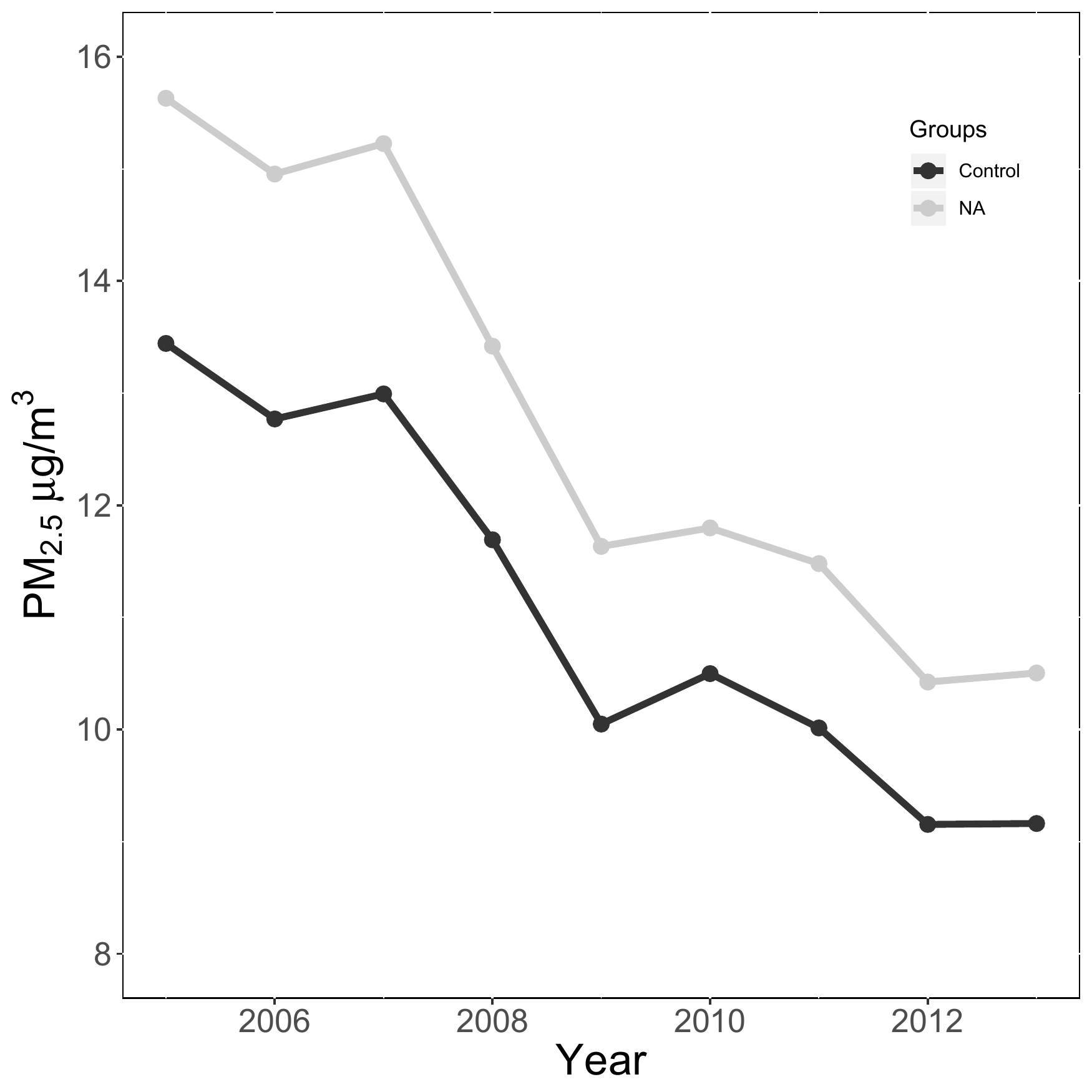}
    \caption{}
  \end{subfigure}
  \begin{subfigure}{6cm}
    \centering\includegraphics[width=6cm]{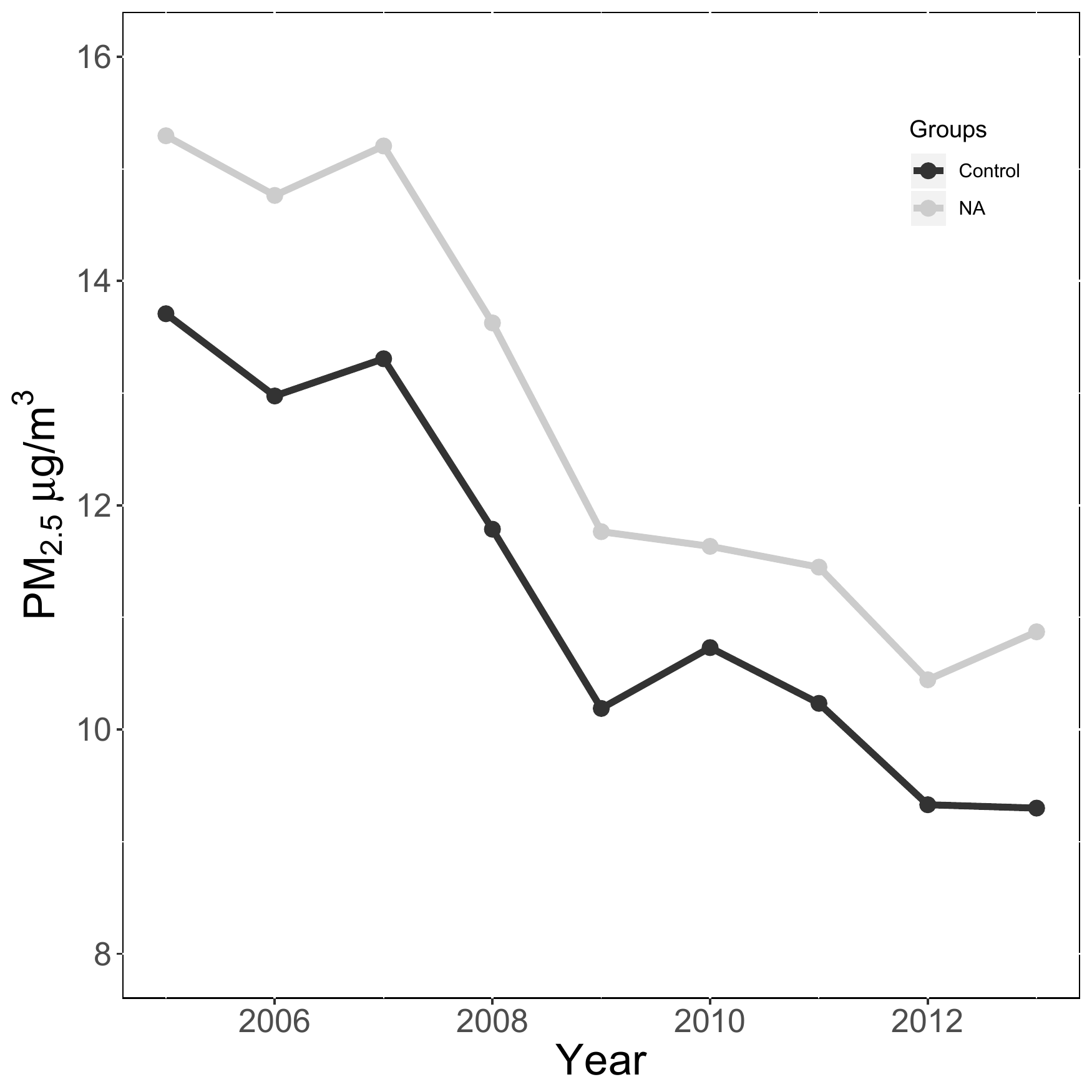}
    \caption{}
  \end{subfigure}
 
  \begin{subfigure}{6cm}
    \centering\includegraphics[width=6cm]{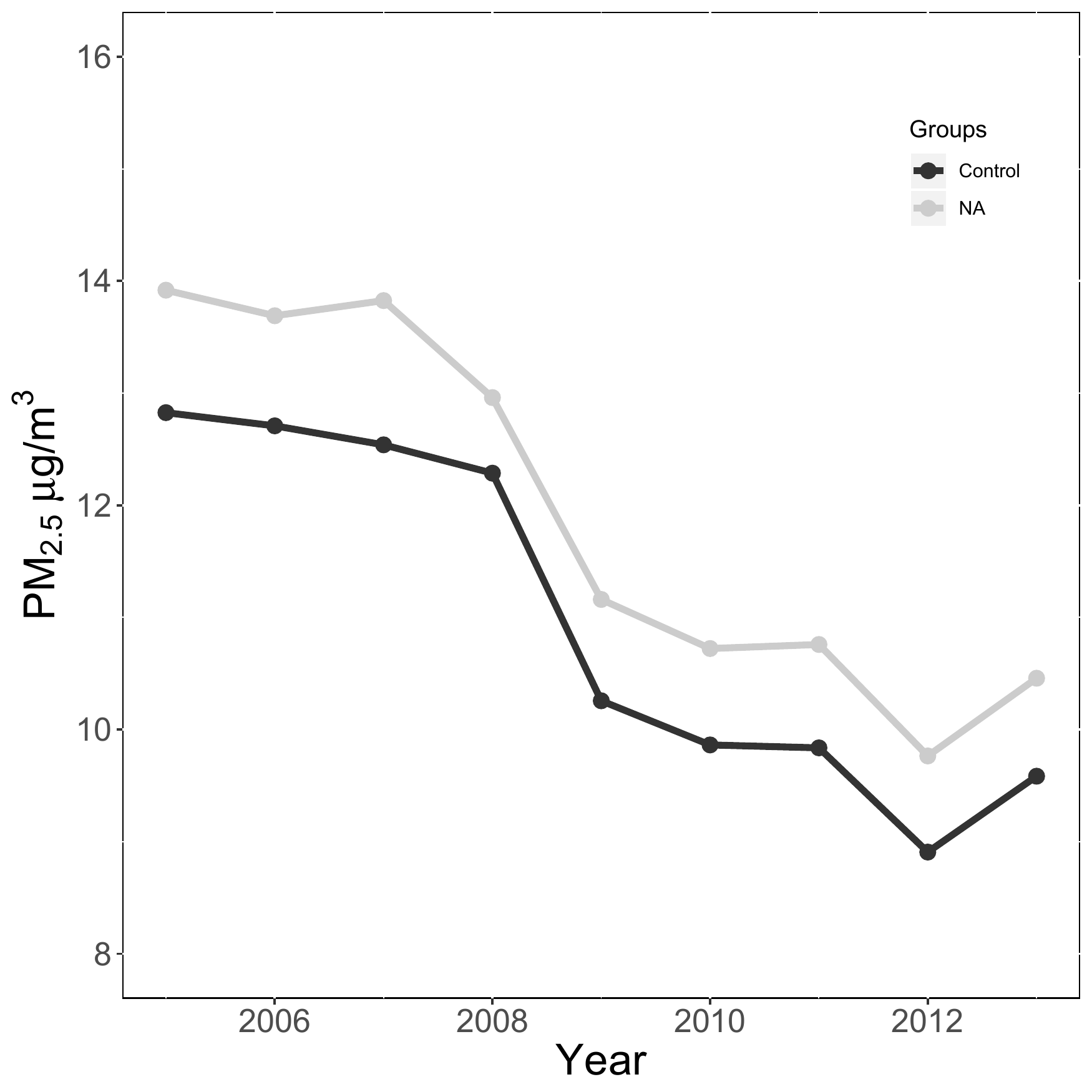}
    \caption{}
  \end{subfigure}
  \begin{subfigure}{6cm}
    \centering\includegraphics[width=6cm]{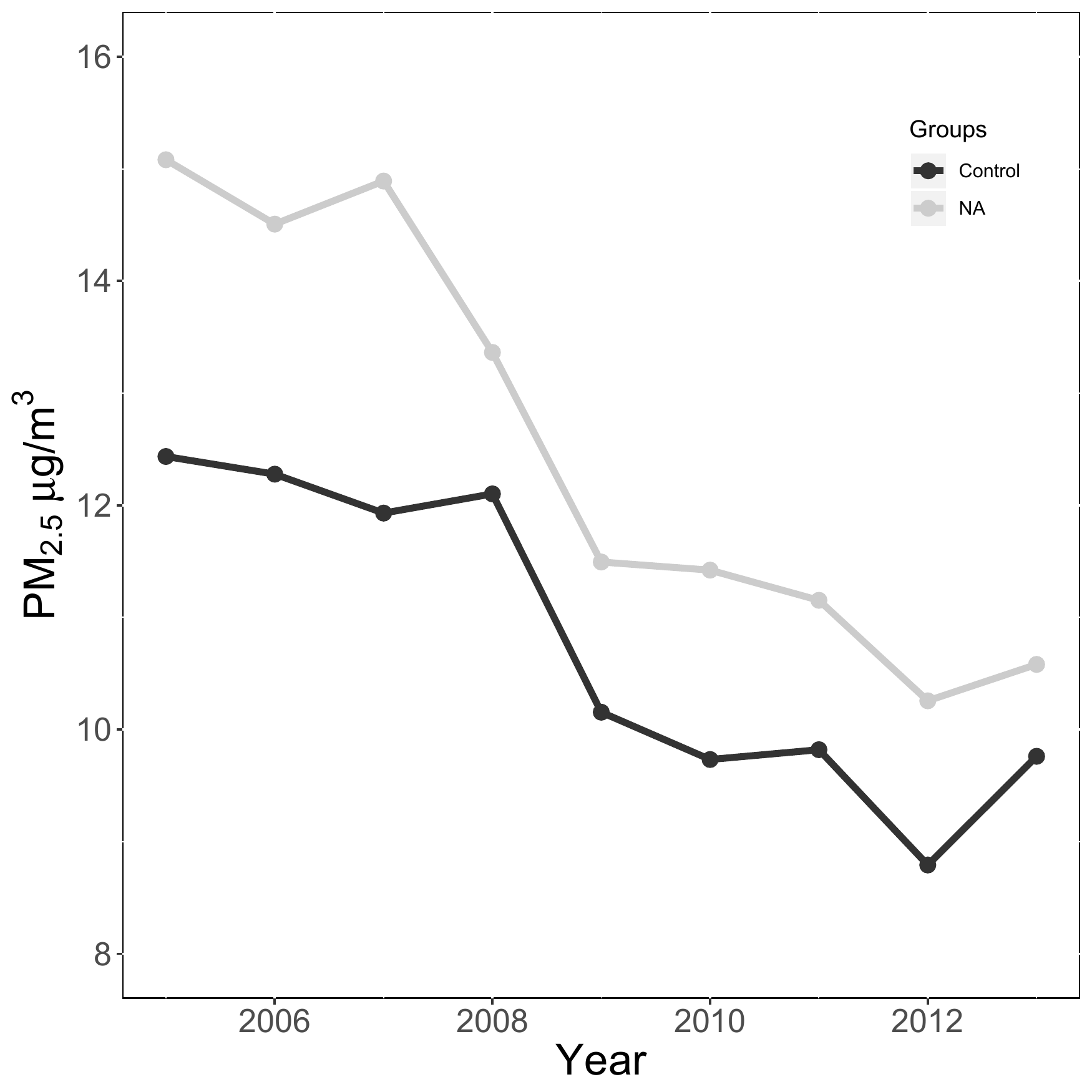}
  \caption{}
  \end{subfigure}
\caption[PM$_{2.5}$ levels for analysis datasets.]{PM$_{2.5}$ levels ($\mu$g/$m^3$) for nonattainment (gray) and control (black) counties across the study period in the a.) full dataset ($N=337$); b.) the pruned dataset ($N=239$); c.) the matched dataset with caliper 0.25 standard deviation ($N=30$); and d.) the matched dataset with caliper 1 standard deviation ($N=36$).}\label{fig:pm25}
\end{figure}

We experienced some of these challenges in the applied dataset. Overall, we found evidence of protective direct and spillover effects from nonattainment designation at the county level. This suggests that county-level actions to reduce PM$_{2.5}$ had a causal effect on lung cancer incidence, and further policy and action should be considered either in these already designated areas or in other areas with elevated lung cancer risk. It is important to note that the true causal effects may not be protective if we have not addressed all confounders, i.e., violated our ignorability under interference assumption. More research needs to be conducted to validate these findings. We also recognize that lung cancer can have an extended latency, and we may see stronger evidence if this analysis considered exposure periods of a decade or longer \cite{HartJaimeE.2014ICES}. We theorize that the spillover effect may be stronger than the direct effect because counties receiving spillover from the nonattainment policy may have a history of lower PM$_{2.5}$ levels (if they are not nonattainment counties themselves) and are benefitting from county-level actions taken to reduce PM$_{2.5}$. The non-significant direct effects may be attributed to the fact that PM$_{2.5}$ levels decreased in both control and nonattainment counties at nearly identical rates across the study period (Figure~\ref{fig:pm25}). This fact remained true for all the study populations. We especially note the uptick in PM$_{2.5}$ from 2012 to 2013, most notable in the matched datasets. Possible explanations include PM$_{2.5}$-reduction actions tapering off with time and PM$_{2.5}$ output sources relocating to other counties not being actively regulated.

We additionally recognize that we made a structural assumption about the spillover from nonattainment counties into other counties. While we explored other spillover structures, we remained surprised that the proportion of nonattainment counties surrounding a county explained the most variation in lung cancer risk.  Spatial analysis in meteorology remains its own challenge. Further work may be done to better model wind direction and PM$_{2.5}$ trajectories, which could enhance spillover estimation in this application\cite{GumiauxC2003Gatb}.

In summary, the methods outlined in this paper motivate further research in air pollution epidemiology. We are especially hopeful to apply these methods to estimate effects of air pollution in China on the United States' West Coast. However, we also believe these methods may be applied beyond air pollution studies to a broader class of observational studies with interference. Researchers increasingly ask questions that involve spatially-related units. While nuances and expert knowledge change from study to study, we believe assumptions and analytical approaches may cater to specific problems and answer causal questions where we can model the relationship between units and the effect of a treatment or an exposure on an outcome.

{\em The data that support the findings of this study are available from the corre- sponding author upon reasonable request.}

\newpage
\section{References}
\bibliography{thesisrefs}

\begin{thebibliography}{10}
\expandafter\ifx\csname url\endcsname\relax
  \def\url#1{\texttt{#1}}\fi
\expandafter\ifx\csname urlprefix\endcsname\relax\def\urlprefix{URL }\fi
\expandafter\ifx\csname href\endcsname\relax
  \def\href#1#2{#2} \def\path#1{#1}\fi

\bibitem{Hernan06}
M.~A. Hern\'{a}n, J.~M. Robins, Estimating causal effects from epidemiologic
  data, Journal of Epidemiologic Community Health 60~(7) (2006) 578--586.

\bibitem{Rubin74}
D.~Rubin, Estimating causal effects of treatments in randomized and
  nonrandomized studies, Journal of Educational Psychology 66 (1974) 688.

\bibitem{Cox58}
D.~Cox, Planning {E}xperiments, New York, NY: Wiley, 1958.

\bibitem{RubinDonaldB.1984BJaR}
D.~B. Rubin, Bayesianly justifiable and relevant frequency calculations for the
  applies statistician, The Annals of Statistics 12~(4) (1984) 1151--1172.

\bibitem{Halloran95}
M.~E. Halloran, C.~J. Struchiner, Causal {I}nference in {I}nfectious
  {D}iseases, Epidemiology 6~(2) (1995) 142--151.

\bibitem{Kao17}
E.~K. Kao, Causal {I}nference {U}nder {N}etworl {I}nterference: {A} {F}ramework
  for {E}xperiments on {S}ocial {N}etworks, Dissertation, Department of
  Statistics, Harvard University.

\bibitem{Toulis12}
P.~Toulis, E.~Kao, Estimation of {C}ausal {P}eer {I}nfluence {E}ffects,
  Proceedings of the 30th International Conference on International Conference
  on Machine Learning 28 (2013) 1489--1497.

\bibitem{Hong06}
G.~Hong, S.~W. Raudenbush, Evaluating {K}indergarten {R}etention {P}olicy,
  Journal of the American Statistical Association 101~(475) (2006) 901--910.

\bibitem{Verbitsky12}
N.~Verbitsky-Savitz, S.~Raudenbush, Causal {I}nference {U}nder {I}nterference
  in {S}patial {S}ettings: {A} {C}ase {S}tudy {E}valuating {C}ommunity
  {P}olicing {P}rogram in {C}hicago, Epidemiologic Methods 1~(1) (2012)
  107--130.

\bibitem{Zigler12}
C.~Zigler, F.~Dominici, Y.~Wang, Estimating causal effects of air quality
  regulations using principal stratification for spatially correlated
  mutlivariate intermediate outcomes, Biostatistics 13~(2) (2012) 289--302.

\bibitem{VanderWeele12}
E.~J.~T. Tchetgen, T.~J. VanderWeele, On causal inference in the presence of
  interference, Statistical Methods in Medical Research 21~(1) (2010) 55--75.

\bibitem{Sobel06}
M.~E. Sobel, What {D}o {R}andomized {S}tudies of {H}ousing {M}obility
  {D}emonstrate?: {C}ausal {I}nference in the {F}ace of {I}nterference, Journal
  of the American Statistical Association 101~(476) (2006) 1398--1407.

\bibitem{Hudgens08}
M.~G. Hudgens, M.~E. Halloran, Toward {C}ausal {I}nference with {I}nterference,
  Journal of the American Statistical Association 103~(482) (2008) 832--842.

\bibitem{Zigler17}
C.~M. Zigler, C.~Choirat, F.~Dominici, Impact of {N}ational {A}mbient {A}ir
  {Q}uality {S}tandards nonattainment designations on particulate pollution and
  health, Epidemiology (2017) 1--30.

\bibitem{Zigler14}
C.~M. Zigler, F.~Dominici, Point: {C}larifying {P}olicy {E}vidence {W}ith
  {P}otential-{O}utcomes {T}hinking\textemdash{B}eyond {E}xposure-{R}esponse
  {E}stimation in {A}ir {P}ollution {E}pidemiology, American Journal of
  Epidemiology 180~(12).

\bibitem{Sussman17}
D.~L. {Sussman}, E.~M. {Airoldi}, {Elements of estimation theory for causal
  effects in the presence of network interference}, ArXiv e-prints\href
  {http://arxiv.org/abs/1702.03578} {\path{arXiv:1702.03578}}.

\bibitem{Athey15}
S.~{Athey}, D.~{Eckles}, G.~{Imbens}, {Exact P-values for Network
  Interference}, ArXiv e-prints\href {http://arxiv.org/abs/1506.02084}
  {\path{arXiv:1506.02084}}.

\bibitem{Manski13}
C.~Manski, Identification of treatment response with social interactions,
  Econometrics Journal 16~(1).
\newblock \href {http://dx.doi.org/10.1111/j.1368-423X.2012.00368.x}
  {\path{doi:10.1111/j.1368-423X.2012.00368.x}}.

\bibitem{ImbensGuido2015}
G.~Imbens, Causal inference for statistics, social, and biomedical sciences :
  an introduction, 2015.

\bibitem{Rubin78}
D.~B. Rubin, Bayesian {I}nference for {C}ausal {E}ffects: {T}he {R}ole of
  {R}andomization, The Annals of Statistics 6 (1978) 34--58.

\bibitem{Rosenbaum83}
P.~H. Rosenbaum, D.~B. Rubin, The {C}entral {R}ole of the {P}ropensity {S}core
  in {O}bservational {S}tudies for {C}ausal {E}ffects, Biometrika 70 (1983)
  41--55.

\bibitem{Papadogeorgou18}
G.~Papadogeorgou, C.~Choirat, C.~M. Zigler, Adjusting for unmeasured spatial
  confounding with distance adjusted propensity score matching, Biostatistics.

\bibitem{Stuart10}
E.~A. Stuart, Matching {M}ethods for {C}ausal {I}nference: {A} {R}eview and a
  {L}ook {F}orward, Statistical Science: a Review Journal of the Institute of
  Mathematical Statistics 25~(1) (2010) 1--21.

\bibitem{R}
{R Core Team}, \href{https://www.R-project.org/}{R: A Language and Environment
  for Statistical Computing}, R Foundation for Statistical Computing, Vienna,
  Austria (2017).
\newline\urlprefix\url{https://www.R-project.org/}

\bibitem{distr}
P.~Ruckdeschel, M.~Kohl, \href{http://www.jstatsoft.org/v59/i04/}{General
  purpose convolution algorithm in {S}4 classes by means of fft}, Journal of
  Statistical Software 59~(4) (2014) 1--25.
\newline\urlprefix\url{http://www.jstatsoft.org/v59/i04/}

\bibitem{overlapping}
M.~Pastore, \href{https://CRAN.R-project.org/package=overlapping}{overlapping:
  Estimation of Overlapping in Empirical Distributions}, r package version
  1.5.0 (2017).
\newline\urlprefix\url{https://CRAN.R-project.org/package=overlapping}

\bibitem{Ord81}
A.~D. Cliff, J.~K. Ord, Spatial {P}rocesses, Pion, 1981.

\bibitem{spdep}
R.~Bivand, G.~Piras, \href{https://www.jstatsoft.org/v63/i18/}{Comparing
  implementations of estimation methods for spatial econometrics}, Journal of
  Statistical Software 63~(18) (2015) 1--36.
\newline\urlprefix\url{https://www.jstatsoft.org/v63/i18/}

\bibitem{Austin11}
P.~C. Austin, An {I}ntroduction to {P}ropensity {S}core {M}ethods for
  {R}educing {E}ffects of {C}onfounding in {O}bservational {S}tudies,
  Multivariate Behavioral Research 46 (2011) 399--424.

\bibitem{doi:10.1002/sim.3680}
D.~Lunn, D.~Spiegelhalter, A.~Thomas, N.~Best,
  \href{https://onlinelibrary.wiley.com/doi/abs/10.1002/sim.3680}{The bugs
  project: Evolution, critique and future directions}, Statistics in Medicine
  28~(25)  3049--3067.
\newblock \href
  {http://arxiv.org/abs/https://onlinelibrary.wiley.com/doi/pdf/10.1002/sim.3680}
  {\path{arXiv:https://onlinelibrary.wiley.com/doi/pdf/10.1002/sim.3680}},
  \href {http://dx.doi.org/10.1002/sim.3680} {\path{doi:10.1002/sim.3680}}.
\newline\urlprefix\url{https://onlinelibrary.wiley.com/doi/abs/10.1002/sim.3680}

\bibitem{gelman1992}
A.~Gelman, D.~B. Rubin, \href{https://doi.org/10.1214/ss/1177011136}{Inference
  from iterative simulation using multiple sequences}, Statist. Sci. 7~(4)
  (1992) 457--472.
\newblock \href {http://dx.doi.org/10.1214/ss/1177011136}
  {\path{doi:10.1214/ss/1177011136}}.
\newline\urlprefix\url{https://doi.org/10.1214/ss/1177011136}

\bibitem{Besag1991}
J.~Besag, J.~York, A.~Molli{\'e},
  \href{https://doi.org/10.1007/BF00116466}{Bayesian image restoration, with
  two applications in spatial statistics}, Annals of the Institute of
  Statistical Mathematics 43~(1) (1991) 1--20.
\newblock \href {http://dx.doi.org/10.1007/BF00116466}
  {\path{doi:10.1007/BF00116466}}.
\newline\urlprefix\url{https://doi.org/10.1007/BF00116466}

\bibitem{Cahoon15}
E.~K. Cahoon, R.~M. Pfeiffer, D.~C. Wheeler, J.~Arhancet, S.-W. Lin, B.~H.
  Alexander, M.~S. Linet, D.~M. Freedman, Relationship between ambient
  ultraviolet radiation and non-{H}odgkin lymphoma subtypes: {A} {U}.{S}.
  population-based study of racial and ethnic groups, International Journal of
  Cancer 136 (2015) 432--441.

\bibitem{EPA16}
U.~S. {E}nvironmental~{P}rotection {A}gency, Summary of the {C}lean {A}ir
  {A}ctHttps://www.epa.gov/laws-regulations/summary-clean-air-act.

\bibitem{EPA16b}
U.~S. {E}nvironmental~{P}rotection {A}gency, Green {B}ook {P}{M}-2.5 (1997)
  {A}rea
  {I}nformationHttps://www.epa.gov/green-book/green-book-pm-25-1997-area-information.

\bibitem{EPACh5}
U.~S. {E}nvironmental~{P}rotection {A}gency, Technical support for state and
  tribal air quality fine particle (pm2.5) designations (2004) 5--1--5--3.

\bibitem{GharibvandLida2017TAbA}
L.~Gharibvand, D.~Shavlik, M.~Ghamsary, W.~L. Beeson, S.~Soret, R.~Knutsen,
  S.~F. Knutsen, The association between ambient fine particulate air pollution
  and lung cancer incidence: Results from the ahsmog-2 study, Environmental
  health perspectives 125~(3).

\bibitem{VilleneuvePaulJ.2014VeaR}
P.~J. Villeneuve, M.~Jerrett, D.~Brenner, J.~Su, H.~Chen, J.~R. McLaughlin,
  Villeneuve et al. respond to ``impact of air pollution on lung cancer'',
  American Journal of Epidemiology 179~(4) (2014) 455--456.

\bibitem{HartJaimeE.2014ICES}
J.~E. Hart, Invited commentary: Epidemiologic studies of the impact of air
  pollution on lung cancer, American Journal of Epidemiology 179~(4) (2014)
  452--454.

\bibitem{Downscaler}
D.~Holland.
\newblock
  \href{https://www.epa.gov/air-research/downscaler-model-predicting-daily-air-pollution}{Downscaler
  model for predicting daily air pollution} [online] (March 2019) [cited March
  22, 2019].

\bibitem{Dwyer-Lindgren2014}
L.~Dwyer-Lindgren, A.~H. Mokdad, T.~Srebotnjak, A.~D. Flaxman, G.~M. Hansen,
  C.~J. Murray, \href{https://doi.org/10.1186/1478-7954-12-5}{Cigarette smoking
  prevalence in us counties: 1996-2012}, Population Health Metrics 12~(1)
  (2014) 5.
\newblock \href {http://dx.doi.org/10.1186/1478-7954-12-5}
  {\path{doi:10.1186/1478-7954-12-5}}.
\newline\urlprefix\url{https://doi.org/10.1186/1478-7954-12-5}

\bibitem{GumiauxC2003Gatb}
C.~Gumiaux, D.~Gapais, J.~Brun, Geostatistics applied to best-fit interpolation
  of orientation data, Tectonophysics 376~(3) (2003) 241--259.

\bibitem{doi:10.1111/1467-9868.00353}
D.~J. Spiegelhalter, N.~G. Best, B.~P. Carlin, A.~Van Der~Linde,
  \href{https://rss.onlinelibrary.wiley.com/doi/abs/10.1111/1467-9868.00353}{Bayesian
  measures of model complexity and fit}, Journal of the Royal Statistical
  Society: Series B (Statistical Methodology) 64~(4)  583--639.
\newblock \href
  {http://arxiv.org/abs/https://rss.onlinelibrary.wiley.com/doi/pdf/10.1111/1467-9868.00353}
  {\path{arXiv:https://rss.onlinelibrary.wiley.com/doi/pdf/10.1111/1467-9868.00353}},
  \href {http://dx.doi.org/10.1111/1467-9868.00353}
  {\path{doi:10.1111/1467-9868.00353}}.
\newline\urlprefix\url{https://rss.onlinelibrary.wiley.com/doi/abs/10.1111/1467-9868.00353}

\bibitem{HodgesReich10}
J.~S. Hodges, B.~J. Reich, Adding Spatially-Correlated Errors Can Mess Up the
  Fixed Effect You Love, Vol.~64, American Statistical Association, 2010.

\end{thebibliography}

\end{document}